\documentclass[reqno]{amsart}
\usepackage{amssymb,epsfig,graphics,graphicx,bbm,hyperref,color}
\usepackage{pgf,pgfarrows,pgfnodes,pgfautomata,pgfheaps,pgfshade}

\usepackage{multicol}
\usepackage{enumitem}
\usepackage{tikz-cd}
\usetikzlibrary{matrix,arrows,decorations.pathmorphing,positioning}

\definecolor{gray}{rgb}{0.93,0.93,0.93}
\definecolor{light-gold}{rgb}{0.99,0.97,0.78}

\def\be{\begin{equation}}
\def\ee{\end{equation}}
\def\bm{\begin{multline}}
\def\bfig{\begin{figure}[htb]}
\def\efig{\end{figure}}
\newcommand{\dd}{{\rm d}}
\newcommand{\e}[1]{\,{\rm e}^{#1}\,}
\newcommand{\ii}{{\rm i}}
\def\Tr{{\operatorname{Tr\,}}}

\newcommand{\sumtwo}[2]{\sum_{\substack{#1 \\ #2}}}

\newcommand{\limtwo}[2]{\lim_{\substack{#1 \\ #2}}}

\numberwithin{equation}{section}

\newtheorem{conjecture}{Conjecture}

\newcommand{\eps}{{\varepsilon}}

\newcommand{\caH}{{\mathcal H}}
\newcommand{\caL}{{\mathcal L}}
\newcommand{\caM}{{\mathcal M}}

\newcommand{\caS}{{\mathcal S}}

\newcommand{\bbC}{{\mathbb C}}
\newcommand{\bbE}{{\mathbb E}}
\newcommand{\bbN}{{\mathbb N}}
\newcommand{\bbP}{{\mathbb P}}
\newcommand{\bbR}{{\mathbb R}}
\newcommand{\bbS}{{\mathbb S}}
\newcommand{\bbZ}{{\mathbb Z}}

\newcommand{\frm}{{\mathfrak m}}
\newcommand{\frn}{{\mathfrak n}}

\makeatletter
  \def\tagform@#1{\maketag@@@{\scriptsize{(#1)}\@@italiccorr}}
\makeatother

\renewcommand{\eqref}[1]{(\ref{#1})}

\begin{document}

\noindent
{\hfill\small Published in {\it 6th Warsaw School of Statistical Physics}, B. Cichocki, M. Napi\'orkowski, J. Piasecki, P. Szymczak eds, Warsaw University Press (2017)} \vspace{10mm}

\title{Universal behaviour of 3D loop soup models}

\author{Daniel Ueltschi}
\address{Department of Mathematics, University of Warwick,
Coventry, CV4 7AL, United Kingdom}
\email{daniel@ueltschi.org}


\keywords{Loop soups, quantum Heisenberg models, Poisson-Dirichlet distribution}

\begin{abstract}
These notes describe several loop soup models and their {\it universal behaviour} in dimensions greater or equal to 3. These loop models represent certain classical or quantum statistical mechanical systems. These systems undergo phase transitions that are characterised by changes in the structures of the loops. Namely, long-range order is equivalent to the occurrence of macroscopic loops. There are many such loops, and the joint distribution of their lengths is always given by a {\it Poisson-Dirichlet distribution}.

This distribution concerns random partitions and it is not widely known in statistical physics. We introduce it explicitly, and we explain that it is the invariant measure of a mean-field split-merge process. It is relevant to spatial models because the macroscopic loops are so intertwined that they behave effectively in mean-field fashion. This heuristics can be made exact and it allows to calculate the parameter of the Poisson-Dirichlet distribution. We discuss consequences about symmetry breaking in certain quantum spin systems.
\end{abstract}

\thanks{\copyright{} 2017 by the author. This paper may be reproduced, in its
entirety, for non-commercial purposes.}

\thanks{Notes prepared for the 6th Warsaw School of Statistical Physics, held from 25 June to 2 July 2016 in Sandomierz, Poland.}

\maketitle

\vspace{-10mm}
{\small\tableofcontents}

\section{Introduction}
\label{sec intro}

``Loop soups" has become the generic term for a statistical physical system where objects are one-dimensional closed trajectories living in a higher dimensional space. Loop soup models do not describe physical systems directly; rather, they are mathematical representations of relevant models. Among many examples of loop soup models, let us mention:
\begin{itemize}
\item Feynman's representation of the interacting Bose gas \cite{Fey}.
\item Lattice permutations \cite{Fey,Kik}: This is a rather crude approximation of the previous system, but the model has interesting physical and mathematical aspects.
\item The Symanzik-BFS loop representation of classical O(N) spin models \cite{BFS,FFS}.
\item O(N) loop models, where the Gibbs factor $\e{\sum_{xy} \beta \vec\varphi_x \cdot \vec\varphi_y}$ is replaced by $\prod_{xy} (1 + \beta \vec\varphi_x \cdot \vec\varphi_y)$. This is justified for small $\beta$.
\item T\'oth's representation of the spin $\frac12$ quantum Heisenberg ferromagnet \cite{Toth}, Aizenman and Nachtergaele's representation of the Heisenberg antiferromagnet \cite{AN}, and extensions that include the spin $\frac12$ quantum XY model \cite{Uel1}.
\end{itemize}

We could add many more examples to this list. The goal of these notes is to show that these loop soup models share a {\it universal feature}: In dimension $d\geq3$, there exists a phase with long, macroscopic loops. Further, the joint distribution of the lengths of long loops is always Poisson-Dirichlet. The latter distribution was explicitly introduced by Kingman \cite{Kin}. It describes random partitions in diverse situations such as population genetics \cite{Ewe}, Bayesian statistics \cite{Fer}, combinatorics \cite{VS}, number theory \cite{Ver}, statistical mechanics \cite{DS}, probability theory \cite{GP}, and record statistics \cite{GMS}. As for loop soup models in statistical physics, that possess a spatial structure, the presence of the Poisson-Dirichlet distribution was pointed out recently in \cite{GUW,GLU,Uel1}.

This conjecture, and the heuristics behind it, involves notions borrowed from mathematical biology and probability theory; they are not well-known in theoretical physics. These notes introduce these notions in an essentially self-contained fashion.

We describe several interesting loop models in Section \ref{sec loop models}. The conjecture about the universal behaviour of loop soups is stated in Section \ref{sec universal}; this involves the Poisson-Dirichlet distribution about random partitions, which is introduced in the following Section \ref{sec PD}. In the next two sections we check that the Poisson-Dirichlet distribution is the invariant measure of the split-merge process; for this, we discuss random permutations in Section \ref{sec random permutations} before introducing the split-merge process in Section \ref{sec split-merge}.

It is a remarkable fact that these mean-field models describe spatial systems exactly; the heuristics is explained in Section \ref{sec heuristics}. It is useful in order to understand the mechanisms, and also to learn a way to calculate the parameter of the Poisson-Dirichlet distribution. We conclude by discussing in Section \ref{sec conseq} a useful consequence of this conjecture, namely that it helps to identify the nature of symmetry breaking in certain quantum spin systems.

\section{Loop soup models}
\label{sec loop models}

\subsection{Feynman representation of the Bose gas}

The representation dates back to 1953 and sought to understand Bose-Einstein condensation in interacting systems. It constitutes an interesting loop model, and it also suggests several related models discussed afterwards.

Recall that the integral kernel of an operator $A : L^2(\bbR^d) \to L^2(\bbR^d)$ is a function $\bbR^d \times \bbR^d \to \bbR$ (which we also denote $A$) that is such that for all square-integrable functions $f$, we have
\be
(Af)(x) = \int_{\bbR^d} A(x,y) f(y) \dd y.
\ee
It is well-known that the integral kernel of the exponential of the laplacian, $\e{\frac12 t \Delta}$, is the gaussian function $g_t(x-y)$, where
\be
g_t(x) = \frac1{(2\pi t)^{d/2}} \e{-x^2 / 2t}.
\ee
The Wiener measure $\dd W$ for the Brownian bridges between $x$ and $y$ is a measure on continuous paths $\omega : [0,\beta] \to \bbR^d$ such that $\omega(0) = x$ and $\omega(\beta) = y$. If $f$ is a function that depends on the path at times $0 < t_1 < \dots < t_k < \beta$, we have
\be
\int_{x \mapsto y} f(\omega) \dd W(\omega) = \int_{\bbR^d} \dd x_1 \dots \int_{\bbR^d} \dd x_k \; g_{t_1}(x_1-x) g_{t_2-t_1}(x_2-x_1) \dots g_{\beta-t_k}(y-x_k) \; f(x_1,\dots,x_k).
\ee
Consider now the operator $\e{\frac12 \Delta - U}$, where the function $U : \bbR^d \to \bbR$ acts as a multiplication operator. Using the Trotter product formula, we can show that the integral kernel of this operator is
\be
\e{\beta (\frac12 \Delta - U)}(x,y) = \int_{x \mapsto y} \dd W(\omega) \e{-\int_0^\beta U(\omega(s)) \dd s}.
\ee

\bfig
\pgfimage[height=50mm]{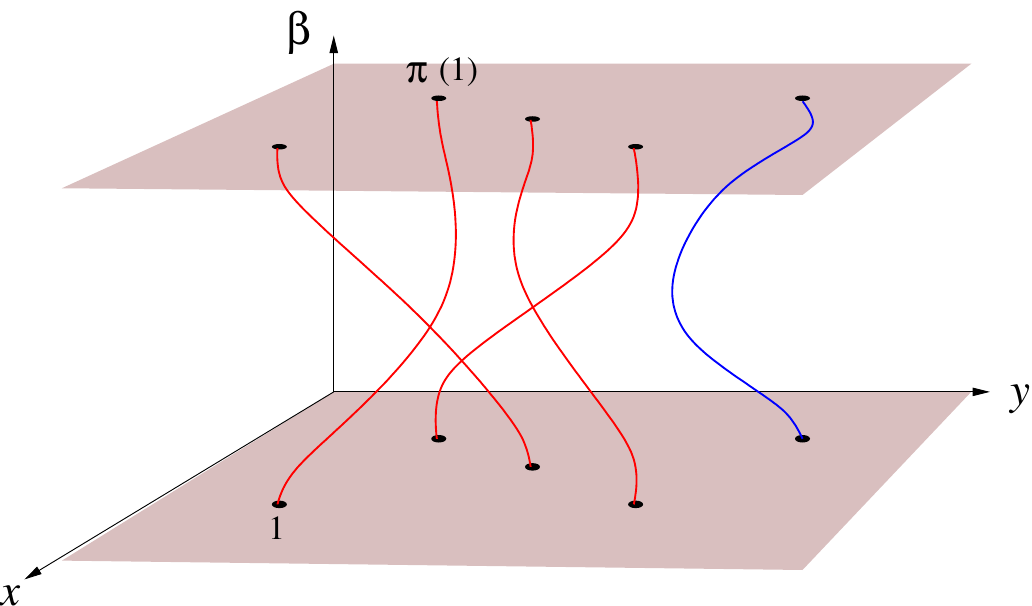}
\caption{Illustration of Feynman's representation of quantum bosons at equilibrium by Brownian trajectories. There are two spatial and one ``imaginary time" dimensions here.}
\label{fig Feynman}
\efig

We now consider a gas of $n$ identical bosons at equilibrium in a domain $\Lambda \subset \bbR^d$, where the two-body interactions between particles are given by the function $U : \bbR^d \to \bbR$. The Hilbert space is the space of square-integrable functions $L^2(\Lambda^n)$ and the hamiltonian is
\be
H_{\Lambda,n} = -\tfrac12 \sum_{i=1}^n \Delta_i + \sum_{1 \leq i<j \leq n} U(x_i-x_j),
\ee
where $\Delta_i$ is the laplacian for the $i$th boson and $U(\cdot)$ acts as multiplication operator. The partition function $Z(\beta,\Lambda,n)$ is given by the trace of $\e{-\beta H_{\Lambda,n}}$ on the symmetric subspace of $L^2(\Lambda^n)$. Let $P_{\rm sym}$ denote the projector onto symmetric functions,
\be
P_{\rm sym} f(x_1,\dots,x_n) = \frac1{n!} \sum_{\sigma \in \caS_n} f(x_{\sigma(1)},\dots,x_{\sigma(n)}).
\ee
The sum is over all permutations of $n$ elements. Then
\be
\begin{split}
Z(\beta,\Lambda,n) &= \Tr_{L^2(\Lambda^n)} P_{\rm sym} \e{-\beta H_{\Lambda,n}} \\
&= \frac1{n!} \sum_{\sigma\in\caS_n} \int_\Lambda \dd x_1 \dots \int_\Lambda \dd x_n \int_{x_1 \mapsto x_{\sigma(1)}} \dd W(\omega_1) \dots \int_{x_n \mapsto x_{\sigma(n)}} \dd W(\omega_n) \\
&\hspace{3cm} \exp \Bigl\{ -\sum_{1\leq i<j \leq n} \int_0^\beta U \bigl( \omega_i(s) - \omega_j(s) \bigr) \dd s \Bigr\}.
\end{split}
\ee
The expression above is illustrated in Fig.\ \ref{fig Feynman}. We observe that it involves a sum over permutations with positive weights; this induces a probability measure on permutations.

One expects that Bose-Einstein condensation is signalled by the occurrence of permutation cycles of divergent lengths (``divergent" refers to the thermodynamic limit where $|\Lambda|,n \to \infty$ while the density $n/|\Lambda|$ is kept fixed); further, these long cycles are {\it macroscopic}, that is, they are proportional to $n$, and there are many of them. This was pointed out by S\"ut\H o in the case of the ideal gas \cite{Suto}. We argue below that this remains true in the presence of interactions, and that the joint distribution of the lengths of macroscopic cycles is Poisson-Dirichlet; this can actually be proved in the case of the ideal gas \cite{BU2}.

\subsection{Lattice permutations}

The model of lattice permutations is more intriguing than physical. It goes back to Feynman \cite{Fey} and Kikuchi \cite{Kik}. It has been studied numerically in \cite{GRU,GLU}, and mathematically in \cite{Betz,BT} --- the latter article proves in particular that the critical parameter for the presence of long cycles is {\it strictly less} than that for self-avoiding walks.

\bfig
\includegraphics[width=60mm]{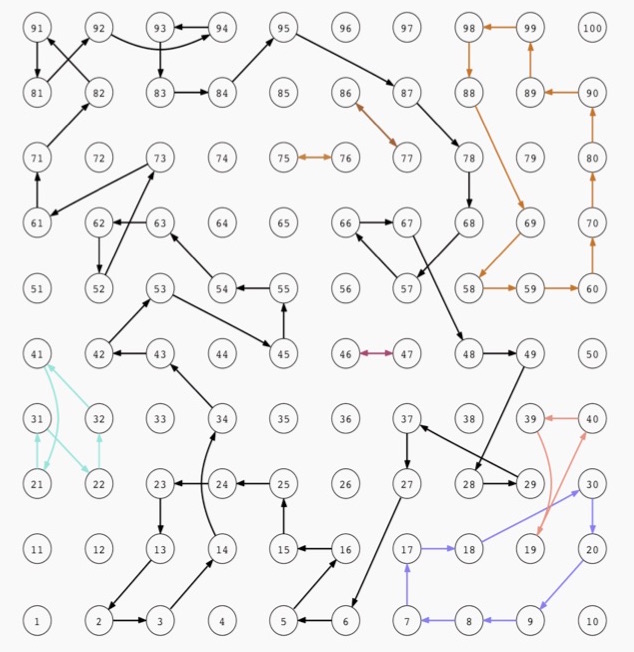}
\caption{Lattice permutations.}
\label{fig lattice permutations}
\efig

Let $\Lambda = \{1,\dots, L\}^d$ be a $d$-dimensional box, and let $\caS_\Lambda$ denote the set of permutations on $\Lambda$ (bijections $\Lambda \to \Lambda$). The probability of the permutation $\sigma \in \caS_\Lambda$ is defined as
\be
\bbP_\Lambda (\sigma) = \frac1{Z(\Lambda)} \exp \Bigl\{ -\alpha \sum_{x\in\Lambda} \xi \bigl( \|x-\sigma(x)\| \bigr) \Bigr\}.
\ee
Here, $\xi$ is an increasing function $[0,\infty) \to [0,\infty]$ such that $\xi(0)=0$, and such that $\e{-\xi(r)}$ decays sufficiently rapidly as $r\to\infty$ so that all jumps $x \mapsto \sigma(x)$ are bounded uniformly in $L$. The normalisation $Z(\Lambda)$ is the partition function
\be
Z(\Lambda) = \sum_{\sigma \in \caS_\Lambda} \exp \Bigl\{ -\alpha \sum_{x\in\Lambda} \xi \bigl( \|x-\sigma(x)\| \bigr) \Bigr\}.
\ee
This model is illustrated in Fig.\ \ref{fig lattice permutations}. It is a simplification of Feynman's representation of the interacting Bose gas; particles are assumed to be spread quite uniformly in the whole domain, hence the lattice. The relevant weight is $\e{-\alpha \|x-\sigma(x)\|^2}$ with $\alpha \sim 1/\beta$; it accounts for the integral over Brownian paths from $x$ to $\sigma(x)$. Interactions between bosons are neglected.

Because of the weights, all jumps $x \mapsto \sigma(x)$ involve nearby sites. The most probable permutation is the identity, $\sigma(x) = x$ for all $x \in \Lambda$. For large $\alpha$, typical permutations are close to the identity with a small density of finite cycles. For small $\alpha$, there are longer jumps, and there is a possibility of very large cycles. A phase transition was indeed observed numerically in \cite{GRU} in dimension $d=3$. Large cycles have {\it macroscopic} lengths, and it was also noticed that the expected length of the longest cycle, divided by the fraction of points in long cycles, was equal to 62\%, as in random permutations without spatial structure. This was a hint pointing to a very general behaviour, but there was no clear understanding then.

The situation has now been clarified. The joint distribution of the lengths of macroscopic cycles is Poisson-Dirichlet, as is explained below. This was numerically verified in this model in \cite{GLU}.

One can also consider an ``annealed" model where one integrates over point positions. Namely, with $\Lambda \subset \bbR^d$ a cubic box of size $L$, the probability of the permutation $\sigma \in \caS_n$ is
\be
\bbP_{\Lambda,n}(\sigma) = \frac1{Z(\Lambda,n)} \int_{\Lambda^n} \dd x_1 \dots \dd x_n \, \exp \Bigl\{ -\alpha \sum_{i=1}^n \xi \bigl( \|x_i - x_{\sigma(i)} \| \bigr) \Bigr\},
\ee
with the normalisation given by
\be
Z(\Lambda,n) = \sum_{\sigma \in \caS_n} \int_{\Lambda^n} \dd x_1 \dots \dd x_n \, \exp \Bigl\{ -\alpha \sum_{i=1}^n \xi \bigl( \|x_i - x_{\sigma(i)} \| \bigr) \Bigr\}.
\ee
This is illustrated in Fig.\ \ref{fig spatial permutations}

\bfig
\pgfimage[height=50mm]{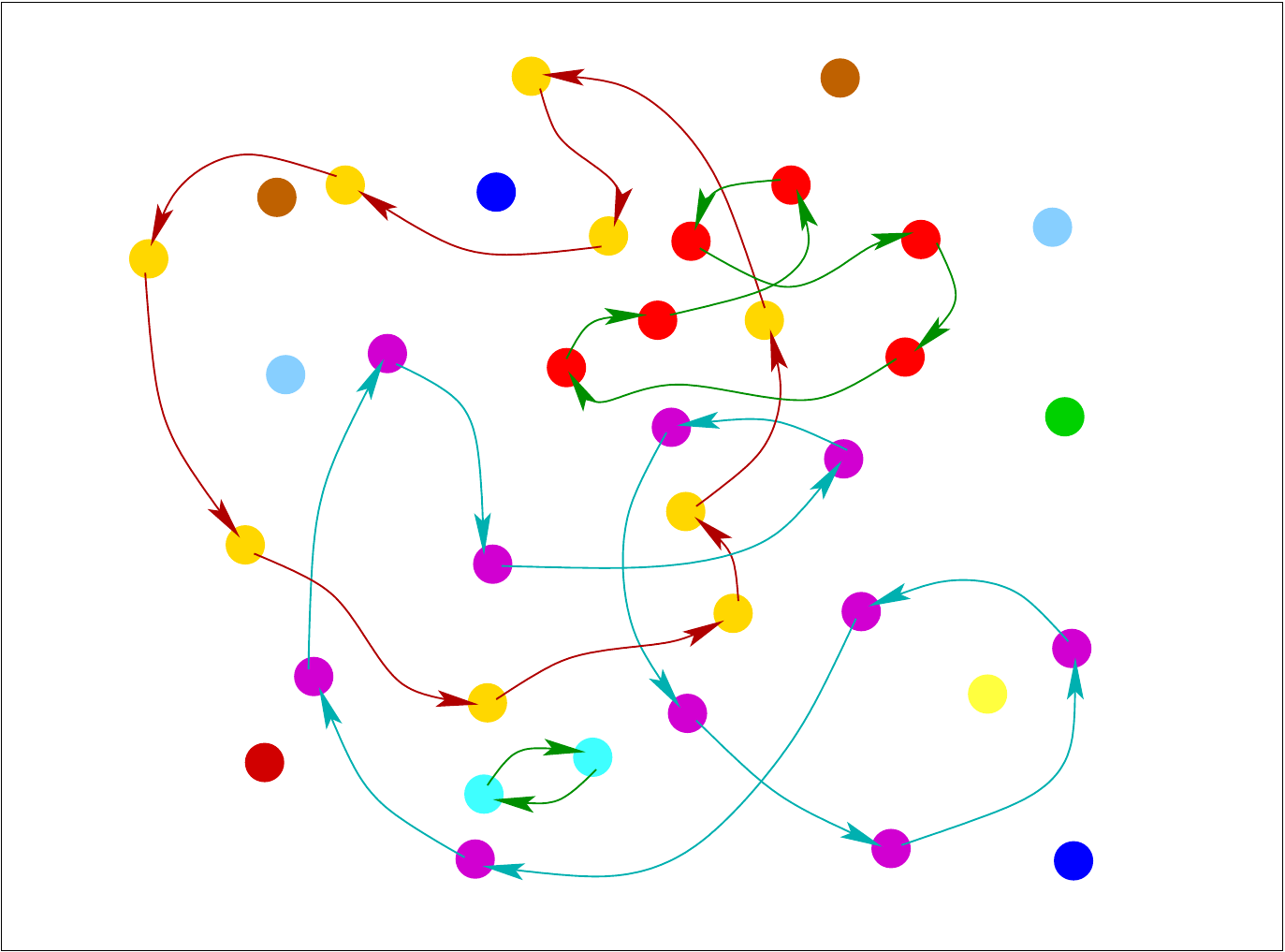}
\caption{Annealed spatial permutations, where one averages over point positions.}
\label{fig spatial permutations}
\efig

The case $\xi(\|x\|) = \|x\|^2$ corresponds to the ideal Bose gas. In this case, S\"ut\H o proved that the Bose-Einstein condensation amounts to the occurrence of macroscopic cycles \cite{Suto}. This was extended in \cite{BU1} to more general functions $\xi$ (such that $\e{-\xi}$ has positive Fourier transform), and the presence of the Poisson-Dirichlet distribution was rigorously established in \cite{BU2}.

\subsection{Spin $O(N)$ models}

Loop representations for classical lattice spin $O(N)$ models were proposed by Brydges, Fr\"ohlich, and Spencer \cite{BFS}; they were partly motivated by earlier work of Symanzik. This representation has allowed to prove the ``triviality" of the behaviour of correlation functions in high dimensions, see \cite{FFS}.

The configuration space is $(\bbS^N)^\Lambda$, where $\bbS^N$ is the $N$-dimensional unit sphere, that is, the set of vectors with $(N+1)$ components and norm 1; the domain $\Lambda$ is a finite subset of $\bbZ^d$. The partition function is
\be
\label{part fct O(N)}
Z(\Lambda) = \int_{(\bbS^N)^\Lambda} \exp\Bigl\{ \tfrac12 \sum_{x,y \in \Lambda, x \neq y} J_{xy} \sigma_x \cdot \sigma_y \Bigr\} \prod_{x\in\Lambda} \dd \sigma_x.
\ee
Here, $(J_{xy})_{x,y\in\Lambda}$ are coupling constants and $\int\dd\sigma_x$ is the Lebesgue integral on $\bbS^N$. The cases $N=1,2,3$ correspond to the Ising model, to the classical XY or rotator model, and to the classical Heisenberg model, respectively.

This partition function can be expressed as a gas of closed loops. Here, a loop of length $k$ is a vector $\gamma = (x_1,\dots, x_k)$ with $x_i \in \Lambda$ and $x_i \neq x_{i+1}$ for $i = 1,\dots,k$ (we identify $x_{k+1}$ with $x_1$). Let $\Gamma(\Lambda)$ denote the set of loops in $\Lambda$, and define the weight $w(\gamma)$ of the loop $\gamma$ by
\be
w(\gamma) = \frac1{2k} \prod_{i=1}^k J_{x_i x_{i+1}}.
\ee
Interactions between loops take a rather simple form; they only depend on the ``local times" $n_x(\cdot)$, $x \in \Lambda$; these local times are given for one or many loops by
\be
\begin{split}
&n_x(\gamma) = \#\{ i=1,\dots,k : x_i = x \}, \\
&n_x(\gamma_1,\dots,\gamma_n) = \sum_{i=1}^n n_x(\gamma_i).
\end{split}
\ee
Let $V : \bbN \to \bbR$ be the function that satisfies
\be
\e{-V(n)} = \frac{\Gamma(\frac N2)}{\Gamma(\frac N2 + n)} \Bigl( \frac N2 \Bigr)^n.
\ee
Notice that $V(0) = V(1) = 0$, and that $V$ is increasing otherwise. The partition function \eqref{part fct O(N)} is then equal to
\bm
Z(\Lambda) = C(N)^{|\Lambda|} \e{-\frac12 \sum_{x,y \in \Lambda} J_{xy}} \sum_{n\geq0} \frac{N^n}{n!} \\
\sum_{\gamma_1, \dots, \gamma_n \in \Gamma(\Lambda)} w(\gamma_1) \dots w(\gamma_n) \exp \Bigl\{ -\sum_{x\in\Lambda} V \bigl( n_x(\gamma_1,\dots,\gamma_n) \bigr) \Bigr\}.
\end{multline}
The constant above is equal to $C(N) = 2 \pi^{N/2+1} / \Gamma(N/2)$ but it is not important. This is indeed a gas of closed loops with ``activity" $w(\gamma)$ and with local interactions. The correlation functions of the original spin model can be expressed in terms of open paths and closed loops. The derivation of this representation is not straightforward and we refer to \cite{BFS,FFS} for two different methods. An amusing remark is that the loop model is well-defined for all $N \in \bbR_+$; in the limit $N \searrow 0$, correlations are given by self-avoiding walks.

Loop $O(N)$ models are simplified models where the weights pick up a factor $N$, and the interactions are local and hard-core. On graphs (lattices) with degree 3, loop $O(N)$ models correspond to a spin model where the Gibbs factor has been approximated,
\be
\e{\frac12 \sum_{x,y} J_{xy} \sigma_x \cdot \sigma_y} \approx \prod_{x,y \in \Lambda} \bigl( 1 + \tfrac12 J_{xy} \sigma_x \cdot \sigma_y \bigr).
\ee
See \cite{NCSOS} for context and definitions, and for a discussion of the joint distribution of the lengths of long loops.

\subsection{Quantum Heisenberg models}
\label{sec Heisenberg}

Some quantum spin systems have loop representations with positive weights. We describe here the loop representations that were progressively introduced in \cite{Toth,AN,Uel1}. Let $\Lambda$ denote the lattice, that is, a finite subset of $\bbZ^d$. The Hilbert space is
\be
\caH_\Lambda = \bigotimes_{x\in\Lambda} \bbC^{2S+1},
\ee
where $S \in \frac12 \bbN$. We consider somewhat artificial pair interactions given by the self-adjoint operators $T_{x,y}, P_{x,y}$, and $Q_{x,y}$, where $x,y \in \Lambda$ are nearest-neighbours; we give below their more familiar expressions in terms of spin operators. These are operators on $\bbC^{2S+1} \otimes \bbC^{2S+1}$ defined as follows:
\begin{itemize}
\item $T_{x,y}$ is the transposition operator, $T_{x,y} |\varphi\rangle \otimes |\psi\rangle = |\psi\rangle \otimes |\varphi\rangle$;
\item $P_{x,y}$ is equal to $(2S+1)$ times the projector onto the spin singlet. If $\{ |a\rangle \}$, $a \in \{-S,\dots,S-1,S\}$ denotes a basis of $\bbC^{2S+1}$, then $P_{x,y}$ has matrix elements
\be
\langle a,b | P_{x,y} | c,d \rangle = (-1)^{a-c} \delta_{a,-b} \delta_{c,-d},
\ee
where $a,b,c,d \in \{-S,\dots,S\}$;
\item $Q_{x,y}$ is as $P_{x,y}$ but without the minus signs, namely
\be
\langle a,b | Q_{x,y} | c,d \rangle = \delta_{a,b} \delta_{c,d}.
\ee
\end{itemize}

The families of hamiltonians involve the parameter $u \in [0,1]$ and are given by
\be
\begin{split}
&H_\Lambda^{(u)} = -\sumtwo{\{x,y\} \subset \Lambda}{\|x-y\|=1} \bigl( uT_{x,y} + (1-u) Q_{x,y} - 1 \bigr), \\
&\tilde H_\Lambda^{(u)} = -\sumtwo{\{x,y\} \subset \Lambda}{\|x-y\|=1} \bigl( uT_{x,y} + (1-u) P_{x,y} - 1 \bigr).
\end{split}
\ee

Let $S_x^i$ denote the $i$th spin operator at site $x$; here, $i=1,2,3$ and $x\in\Lambda$. In the case $S=\frac12$, the first hamiltonian is
\be
\label{ham spin 12}
H_\Lambda^{(u)} = -2 \sumtwo{\{x,y\} \subset \Lambda}{\|x-y\|=1} \bigl( S_x^1 S_y^1 + (2u-1) S_x^2 S_y^2 + S_x^3 S_y^3 - \tfrac14 \bigr).
\ee
We get the usual spin $\frac12$ Heisenberg ferromagnet with $u=1$; the quantum rotator model, or quantum XY model, with $u = \frac12$; and we get a model that is unitarily equivalent to the Heisenberg antiferromagnet with $u=0$.

In the case $S=1$ the second hamiltonian $\tilde H_\Lambda^{(u)}$ is more relevant and is given by
\be
\label{ham spin 1}
\tilde H_\Lambda^{(u)} = -\sumtwo{\{x,y\} \subset \Lambda}{\|x-y\|=1} \bigl( u \vec S_x \cdot \vec S_y + (\vec S_x \cdot \vec S_y)^2 - 2 \bigr).
\ee
We discuss the phase diagram of this model in Section \ref{sec conseq}; as will be explained there, the Poisson-Dirichlet conjecture can be used to identify the nature of extremal states at low temperatures.

We now describe the derivation of the loop model. The partition function can be expanded using the Trotter product formula, which yields a sort of classical model in one more dimension.
Recall that a Poisson point process on the interval $[0,1]$ describes the occurrence of independent events at random times. Let $u\geq0$ be the intensity of the process. The probability that an event occurs in the infinitesimal interval $[t,t+\dd t]$ is $u \dd t$; disjoint intervals are independent. Poisson point processes are relevant to us because of the following expansion of the exponential of matrices:
\be
\exp\Bigl\{ u \sum_{i=1}^{k} (M_{i}-1) \Bigr\} = \int \rho(\dd\omega) \prod_{(i,t) \in \omega} M_{i},
\ee
where $\rho$ is a Poisson point process on $\{1,\dots,k\} \times [0,1]$ with intensity $u$, and the product is over the events of the realisation $\omega$ in increasing times. (To prove it, use the Trotter product formula in the left side so as to get a discretised Poisson process, which converges to the right side.) We actually consider an extension where the time intervals are labeled by the edges of the lattice, and where two kinds of events occur with respective intensities $u$ and $1-u$. Then
\be
\label{Poisson exp}
\exp\Bigl\{ -\sum_{\langle x,y \rangle} \bigl( uM_{xy}^{(1)} + (1-u) M_{xy}^{(2)} - 1 \bigr) \Bigr\} = \int\rho(\dd\omega) \prod_{(x,y,i,t) \in \omega} M_{xy}^{(i)}.
\ee
The product is over the events of $\omega$ in increasing times; the label $i$ is equal to 1 if the event is of the first kind, and 2 if the event is of the second kind.

Let $\sigma = (\sigma_x)_{x\in\Lambda}$, with $\sigma_{x} \in \{-S,\dots,S\}$, be a ``classical spin configuration", and let $|\sigma\rangle = \otimes_{x\in\Lambda} |\sigma_{x}\rangle$ denote the elements of the orthonormal basis of $\caH_{\Lambda}$ where $S_{x}^{3}$ are diagonal. Applying the Poisson expansion \eqref{Poisson exp}, we get
\bm
\label{full expansion}
\Tr \e{-\sum_{\langle x,y \rangle} (uT_{xy} + (1-u) Q_{xy} -1)} = \int\rho(\dd\omega) \sum_{\sigma_{1}, \dots, \sigma_{k}} \\
\langle\sigma_{1}| M_{x_{k} y_{k}}^{(i_{k})} |\sigma_{k}\rangle \langle\sigma_{k}| M_{x_{k-1} y_{k-1}}^{(i_{k-1})} |\sigma_{k-1}\rangle \dots \langle\sigma_{2}| M_{x_{1} y_{1}}^{(i_{1})} |\sigma_{1}\rangle.
\end{multline}
Here, $(x_{1}, y_{1}, i_{1}), \dots, (x_{k},y_{k},i_{k})$ are the events of the realisation $\omega$ in increasing times. The number of events $k$ is random.

\begin{centering}
\bfig
\begin{picture}(0,0)%
\includegraphics{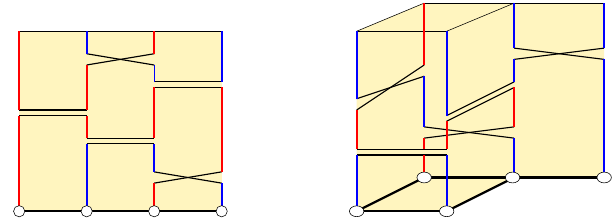}
\end{picture}%
\setlength{\unitlength}{2368sp}%
\begingroup\makeatletter\ifx\SetFigFont\undefined%
\gdef\SetFigFont#1#2#3#4#5{%
  \reset@font\fontsize{#1}{#2pt}%
  \fontfamily{#3}\fontseries{#4}\fontshape{#5}%
  \selectfont}%
\fi\endgroup%
\begin{picture}(8159,2937)(1246,-4090)
\put(5731,-4036){\makebox(0,0)[lb]{\smash{{\SetFigFont{8}{9.6}{\rmdefault}{\mddefault}{\updefault}{\color[rgb]{0,0,0}$0$}%
}}}}
\put(7876,-3961){\makebox(0,0)[lb]{\smash{{\SetFigFont{8}{9.6}{\rmdefault}{\mddefault}{\updefault}{\color[rgb]{0,0,0}$\Lambda$}%
}}}}
\put(4351,-4036){\makebox(0,0)[lb]{\smash{{\SetFigFont{8}{9.6}{\rmdefault}{\mddefault}{\updefault}{\color[rgb]{0,0,0}$\Lambda$}%
}}}}
\put(1276,-1636){\makebox(0,0)[lb]{\smash{{\SetFigFont{8}{9.6}{\rmdefault}{\mddefault}{\updefault}{\color[rgb]{0,0,0}$\beta$}%
}}}}
\put(5776,-1636){\makebox(0,0)[lb]{\smash{{\SetFigFont{8}{9.6}{\rmdefault}{\mddefault}{\updefault}{\color[rgb]{0,0,0}$\beta$}%
}}}}
\put(1246,-4036){\makebox(0,0)[lb]{\smash{{\SetFigFont{8}{9.6}{\rmdefault}{\mddefault}{\updefault}{\color[rgb]{0,0,0}$0$}%
}}}}
\end{picture}%
\caption{Graphs and realisations of Poisson point processes, and their loops. In both cases, the number of loops is $|\caL(\omega)| = 2$.}
\label{fig loops}
\efig
\end{centering}

This expansion has a convenient graphical description. Namely, we view $\rho(\dd\omega)$ as the measure of a Poisson point process for each edge of $\Lambda$, where ``crosses'' occur with intensity $u$ and ``double bars'' occur with intensity $1-u$. In order to find the loop that contains a given point $(x,t) \in \Lambda \times [0,\beta]$, one can start by moving upwards, say, until one meets a cross or a double bar. Then one jumps onto the corresponding neighbour; if the transition is a cross, one continues in the same vertical direction; if it is a double bar, one continues in the opposite direction. The vertical direction has periodic boundary conditions. See Fig.\ \ref{fig loops} for an illustration.

The sum over $|\sigma_{i}\rangle$ is then equivalent to assigning independent labels to each loop. Indeed, in \eqref{full expansion}, the matrix elements of $T_{xy}$ and $Q_{xy}$ force the spin values to stay constant along the loops at each cross and at each double bar. This is illustrated in Fig.\ \ref{fig spinconfig}.

\begin{centering}
\bfig
\includegraphics[width=90mm]{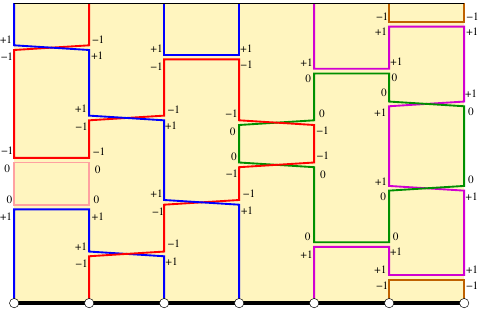}
\caption{Illustration for a realisation of the process $\rho(\dd\omega)$ and a compatible space-time spin configuration. Here, one considers the case $S=1$, where spin values belong to $\{-1,0,1\}$.}
\label{fig spinconfig}
\efig
\end{centering}

We then obtain an expression for the partition function, namely
\be
Z_{\Lambda}^{(u)} = \Tr_{\caH_{\Lambda}} \e{-\beta H_{\Lambda}^{(u)}} = \int\rho(\dd\omega) \sum_{\sigma:\omega} 1 = \int (2S+1)^{|\caL(\omega)|} \rho(\dd\omega).
\ee
The sum in the middle term is over a spin assignment to each loop; there are exactly $2S+1$ possibilities for each loop, hence the result.
Let $\bbP_\Lambda^{(u)}$ denote the probability with respect to the measure $\frac1{Z_{\Lambda}} (2S+1)^{|\caL(\omega)|} \rho(\dd\omega)$. The spin-spin correlation function can be calculated using the same expansion as for the partition function. We get
\be
\label{loop spin correl}
\Tr S_{x}^{3} S_{y}^{3} \e{-\beta H_{\Lambda}^{(u)}} = \int\rho(\dd\omega) \sum_{\sigma:\omega} \sigma_{x,0} \sigma_{y,0}.
\ee
The sum is over all possible labels for the loops, and $\sigma_{x,0}$ denotes the label at site $x$ and time 0. The sum is zero unless $x$ and $y$ belong to the same loop (at time 0), in which case one can check that it gives $\frac13 S (S+1) (2S+1)^{|\caL(\omega)|}$. Then
\be
\langle S_x^3 S_y^3 \rangle = \frac1{Z_{\Lambda}^{(u)}} \Tr S_{x}^{3} S_{y}^{3} \e{-\beta H_{\Lambda}^{(u)}} = \tfrac13 S (S+1) \bbP_\Lambda^{(u)} \bigl( x \leftrightarrow y \bigr).
\ee

\bfig
\includegraphics[width=120mm]{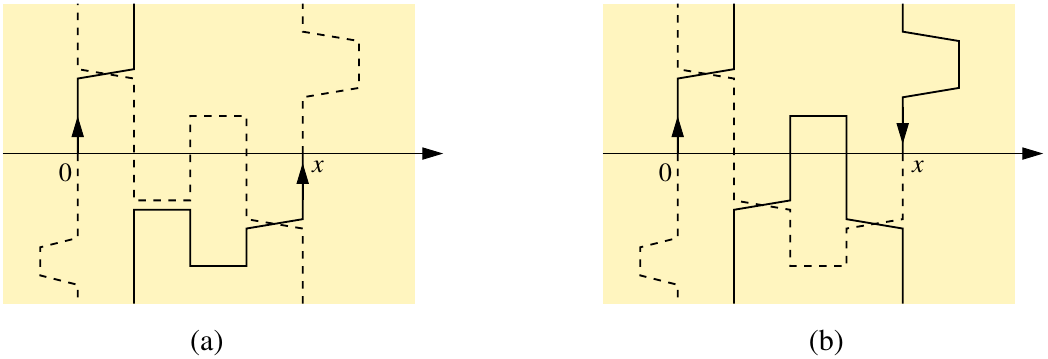}
\caption{Illustration for the two-point correlation function $\langle S_x^2 S_y^2 \rangle$, as expressed in \eqref{interesting spin-spin}.}
\label{fig correl}
\efig

The correlation function $\langle S_x^1 S_y^1 \rangle$ is equal to $\langle S_x^3 S_y^3 \rangle$ by spin symmetry, but correlations $\langle S_x^2 S_y^2 \rangle$ are different. In order to find the loop equivalent for the latter correlation, we write a similar expansion but with additional factors $\langle \sigma_{x,0-} | S_{x}^{2} | \sigma_{x,0+} \rangle$ and $\langle \sigma_{y,0-} | S_{y}^{2} | \sigma_{y,0+} \rangle$. These factors force $(x,0)$ and $(y,0)$ to be in the same loop.  Now recall that $S_x^{2}= \frac1{2\ii} (S_x^{+}-S_x^{-})$ while $S_x^{1} = \frac12 (S_x^{+} + S_x^{-})$. If the loop connection is as in Fig.\ \ref{fig correl} (a), there is one factor with $S^{+}$ and one factor with $S^{-}$ (on either site) , resulting in $-\ii^{2}$ times the same contribution as for $S^{1}$. On the other hand, if the connection is as in Fig.\ \ref{fig correl} (b), both factors involve $S^{+}$ or both involve $S^{-}$, and the contribution is $\ii^{2}$ times that of $S^{1}$. We find
\be
\label{interesting spin-spin}
\begin{tikzpicture}[>=angle 90]
\node (n0) at (0,0) {$\;\;\;\;\; \langle S_{x}^{2} S_{y}^{2} \rangle = \tfrac13 S(S+1) \Bigl[ \bbP_\Lambda^{(u)} \Bigl($};
\node (n1) at (2.4,0) {$x$};
\node (n2) at (3.1,-0.04) {$y$};
\draw[-] (n1) edge[out=90,in=270] (n2);
\draw[-] (n1) edge[out=270,in=90] (n2);
\node (n3) at (3.9,0) {$\Bigr) - \bbP_\Lambda^{(u)} \Bigl($};
\node (n4) at (4.75,0) {$x$};
\node (n5) at (5.35,-0.04) {$y$};
\draw[-] (n4) edge[out=90,in=90] (n5);
\draw[-] (n4) edge[out=270,in=270] (n5);
\node (n6) at (5.7,0) {$\Bigr) \Bigr].$};
\end{tikzpicture}
\ee

The representation for the family with hamiltonian $\tilde H_\Lambda^{(u)}$ is similar, but with a few important differences. Instead of being constant along loops, the spin values change signs at double bars, that is, when the vertical direction of the trajectory changes. The minus signs in the matrix elements of $P_{xy}$ cancel when $S \in \bbN$, but the representation for half-integer spins has unwelcome signs. See \cite{Uel2} for more details.

The model with $u=1$ involves random permutations and is also known as the random interchange model, or random stirring. There exist mathematical studies on the complete graph \cite{Sch, BK, Bjo1, Bjo2} and on the hypercube \cite{KMU}.

\section{Universal behaviour of loop soups}
\label{sec universal}

Consider an arbitrary loop soup model with the following mathematical structure. To each outcome (loop configuration) corresponds a set of $k$ loops ($k$ varies) with lengths $\ell_1,\dots\ell_k$. We assume that loops have been ordered so that $\ell_1 \geq \ell_2 \geq \dots \geq \ell_k$; the loops occupy a domain of volume $V = \sum_{i=1}^k \ell_i$. We let $\bbP_V$ and $\bbE_V$ denote the probability measure and expectation of this loop soup. We also suppose that there is a notion of infinite-volume limit $V\to\infty$. The following vector is a random partition of the interval $[0,1]$:
\be
\Bigl( \frac{\ell_1}V, \frac{\ell_2}V, \dots, \frac{\ell_k}V \Bigr).
\ee
We call a loop {\it macroscopic} if $\ell_i \sim V$, and {\it microscopic} if $\ell_i \sim 1$; it is {\it mesoscopic} otherwise, that is, if $1 \ll \ell_i \ll V$.

There are two conjectures. The first one states that macroscopic loops occupy a fixed portion of the volume, and that microscopic loops occupy the rest; there are certainly mesoscopic loops as well, but they occupy a negligible fraction of the volume. Let us emphasise that this conjecture is expected to be relevant in dimensions 3 and more (and also in the ground state of two-dimensional quantum systems); it is not expected to hold in loop soups of dimensions 1 and 2.

\begin{conjecture}
\label{conj 1}
There exists $m \in [0,1]$ such that for every $\eps>0$:
\[
\boxed{
\begin{split}
&\lim_{n\to\infty} \lim_{V\to\infty} \bbP_V \Bigl( \sum_{i=1}^n \frac{\ell_i}V \in [m-\eps,m+\eps] \Bigr) = 1; \\
&\lim_{n\to\infty} \lim_{V\to\infty} \bbP_V \Bigl( \sum_{i\geq1: \ell_i<n} \frac{\ell_i}V \in [1-m-\eps,1-m+\eps] \Bigr) = 1.
\end{split}
}
\]
\end{conjecture}

It follows from this conjecture that typical partitions have the form displayed in Fig.\ \ref{fig partition}, with $m$ almost always taking the same value.

\begin{centering}
\bfig
\begin{picture}(0,0)%
\includegraphics{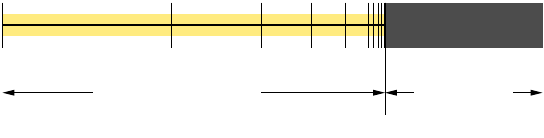}
\end{picture}%
\setlength{\unitlength}{2368sp}%
\begingroup\makeatletter\ifx\SetFigFont\undefined%
\gdef\SetFigFont#1#2#3#4#5{%
  \reset@font\fontsize{#1}{#2pt}%
  \fontfamily{#3}\fontseries{#4}\fontshape{#5}%
  \selectfont}%
\fi\endgroup%
\begin{picture}(7245,1819)(1168,-3058)
\put(2476,-2491){\makebox(0,0)[lb]{\smash{{\SetFigFont{8}{9.6}{\rmdefault}{\mddefault}{\updefault}{\color[rgb]{0,0,0}macroscopic, PD($\vartheta$)}%
}}}}
\put(6766,-2506){\makebox(0,0)[lb]{\smash{{\SetFigFont{8}{9.6}{\rmdefault}{\mddefault}{\updefault}{\color[rgb]{0,0,0}microscopic}%
}}}}
\put(6181,-2986){\makebox(0,0)[lb]{\smash{{\SetFigFont{11}{13.2}{\rmdefault}{\mddefault}{\updefault}{\color[rgb]{0,0,0}$m$}%
}}}}
\end{picture}%
\caption{A typical partition of a loop soup model in dimensions three and higher. The partition in the interval $[0,m]$ follows a Poisson-Dirichlet distribution; the partition in the interval $[m,1]$ consists of microscopic elements. Elements of intermediate size occupy a vanishing interval.}
\label{fig partition}
\efig
\end{centering}

The second conjecture states that the lengths of macroscopic loops are given by a Poisson-Dirichlet distribution for a suitable parameter $\vartheta$. (This family of distributions is introduced in Section \ref{sec PD}.) This conjecture can be stated in different ways, we suggest three of them.

\begin{conjecture}
\label{conj2}
Assume that $m>0$ in Conjecture \ref{conj 1}. Then there is $\vartheta \in (0,\infty)$ such that the following three claims hold true.
\begin{enumerate}
\item For any fixed $n$, the joint distribution of the vector $\bigl( \frac{\ell_1}{mV}, \dots, \frac{\ell_n}{mV} \bigr)$ converges as $V\to\infty$ to the joint distribution of the first $n$ elements of a random partition with {\rm PD}($\vartheta$) distribution.

\item For any $n \in \bbN$ and any $a_1,\dots,a_n > 1$ the moments of $\bigl( \frac{\ell_1}{mV}, \dots, \frac{\ell_n}{mV} \bigr)$ converge as $V\to\infty$ to the moments of {\rm PD}($\vartheta$); precisely,
\[
\boxed{
\lim_{V\to\infty} \bbE_V \biggl( \sumtwo{j_{1}, \dots, j_n \geq 1}{\rm distinct} \Bigl( \frac{\ell_{j_{1}}}{mV} \Bigr)^{a_{1}} \dots \Bigl( \frac{\ell_{j_{n}}}{mV} \Bigr)^{a_n} \biggr) = \frac{\vartheta^{n} \, \Gamma(\vartheta) \, \Gamma(a_{1}) \dots \Gamma(a_{n})}{\Gamma(\vartheta + a_{1} + \dots + a_{n})}.
}
\]

\item Let $f$ be a differentiable function $[0,1] \to \bbR$ such that $f(0)=1$ and $f'(0)=0$. Then
\[
\boxed{
\lim_{V\to\infty} \bbE_V \Bigl( \prod_{j\geq1} f \bigl( \tfrac{\ell_j}V \bigr) \Bigr) = \bbE_{{\rm PD}(\vartheta)} \Bigl( \prod_{j\geq1} f(m X_j) \Bigr).
}
\]
\end{enumerate}
\end{conjecture}

Notice that in part (2), the $a_i$s cannot be less than 1 (the limit would diverge), and cannot be equal to 1 either (the sum $\sum_j \frac{\ell_j}{mV}$ gives $1/m$ instead of 1); with $a_i>1$, the contribution of microscopic loops vanishes in the limit $V\to\infty$. The formula for the moment was derived in \cite{NCSOS} in the context of O(N) loop models using ``supersymmetric" calculations.

In order to understand the part (3) of the conjecture, let us take $f(x) = \e{x^2}$; then
\be
\bbE_V \Bigl( \prod_{j\geq1} f \bigl( \tfrac{\ell_j}V \bigr) \Bigr) = \bbE_V \Bigl( \e{\sum_{j\geq1} (\frac{\ell_j}V)^2} \Bigr).
\ee
The number of microscopic loops is of order $V$ and each contributes $\sim 1/V^2$, so they can be neglected; the expectation picks up macroscopic loops only. This form of the conjecture is very useful for the study of symmetry breaking in quantum spin systems; see Section \ref{sec heuristics}.

As mentioned before, the first hint of a universal behaviour was found in a numerical study of lattice permutations \cite{GRU}. These conjectures were first made in \cite{GUW}. An important article is Schramm's study of the random interchange model on the complete graph \cite{Sch}; it owes much to a heuristics originally proposed by Aldous, based on the split-merge process. There is now much evidence for the validity of Conjectures 1 and 2. This has been established in the annealed model of spatial permutations in a mathematically rigorous fashion \cite{BU2}. It is also backed by numerical studies for the model of lattice permutations \cite{GLU}; for loop $O(N)$ models \cite{NCSOS}; and for the random loop models of Section \ref{sec Heisenberg} \cite{BBBU}.

\section{Random partitions and Poisson-Dirichlet distributions}
\label{sec PD}

The lengths of long loops have the mathematical structure of random partitions. Recall that a partition of the interval $[0,m]$ is a (finite or infinite) sequence of decreasing positive numbers $(\lambda_1,\lambda_2,\dots)$ such that $\sum_{j\geq1} \lambda_j = m$.
We will also consider sequences of positive numbers that are not necessarily decreasing; we still call such a sequence an (unordered) partition.

We review the mathematical notions and relevant properties.

\subsection{Residual allocation, or stick breaking construction}

Let $\nu_1$ be a probability measure on the interval $[0,1]$; we assume that it has a continuous probability density function. For $m>0$, we denote $\nu_m$ the rescaled measure on $[0,m]$, that is, it satisfies $\bbP_{\nu_1}(X<s) = \bbP_{\nu_m}(X<ms)$ for $s \in [0,1]$.

We construct a random sequence of positive numbers $X_1, X_2, \dots$ with the following induction:
\begin{itemize}
\item Choose $X_1$ according to $\nu_1$.
\item Choose $X_2$ according to $\nu_{1-X_1}$; notice that $\frac{X_2}{1-X_1}$ has distribution $\nu_1$.
\item Choose $X_3$ according to $\nu_{1-X_1-X_2}$; notice that $\frac{X_3}{1-X_1-X_2}$ has distribution $\nu_1$.
\item Etc...
\end{itemize}
This gives a sequence of positive numbers $(X_1,X_2,\dots)$ that tends to 0 and such that $\sum_{j\geq1} X_j = 1$. This is an unordered random partition of $[0,1]$.

Let the random numbers $Y_1,Y_2,\dots$ be defined from the $X_i$s by
\be
\label{Y(X)}
\begin{split}
&Y_1 = X_1; \\
&Y_2 = \tfrac{X_2}{1-X_1}; \\
&Y_3 = \tfrac{X_3}{1-X_1-X_2}; \\
&{\rm etc...}
\end{split}
\ee
As noticed above, the $Y_i$s are independent and identically distributed with distribution $\nu_1$. Further, the following equation is easy to verify:
\be
1 - X_1 - ... - X_{k+1} = (1 - X_1 - ... - X_k) \underbrace{\bigl( 1 - \tfrac{X_{k+1}}{1 - X_1 - ... - X_k} \bigr)}_{1-Y_{k+1}}.
\ee
It follows by induction that
\be
1 - X_1 - \dots - X_k = (1-Y_1) \dots (1-Y_k),
\ee
which allows to invert the relations \eqref{Y(X)}
\be
\begin{split}
&X_1 = Y_1; \\
&X_2 = (1-Y_1) Y_2; \\
&X_3 = (1-Y_1) (1-Y_2) Y_3; \\
&{\rm etc...}
\end{split}
\ee

Consider a random partition of $[0,1]$ obtained through the stick breaking construction above, and two random numbers $T,U \in [0,1]$ (independent, uniformly distributed). What is the probability that they fall in the same partition element? This calculation can be performed, and the result turns out to be useful. Recall that the probability of an event is equal to the expectation of the indicator function on this event. Let $\bbP_{{\rm RA}(\nu_1)}$ and $\bbE_{{\rm RA}(\nu_1)}$ denote the probability and expectation of random partitions distributed according to residual allocation with measure $\nu_1$ on $[0,1]$. We have
\be
\begin{split}
\bbP_{{\rm RA}(\nu_1)}(T,U \in \text{ $k$th partition element}) &= \int_0^1 \dd t \int_0^1 \dd u \; \bbE_{{\rm RA}(\nu_1)} \bigr( 1_{t \in X_k} 1_{u \in X_k} \bigr) \\
&= \bbE_{{\rm RA}(\nu_1)}(X_k^2) \\
&= \bbE_{{\rm RA}(\nu_1)} \bigl( (1-Y_1)^2 \dots (1-Y_{k-1})^2 Y_k^2 \bigr) \\
&= \bbE_{\nu_1} \bigl( (1-Y)^2 \bigr)^{k-1} \bbE_{\nu_1}(Y^2).
\end{split}
\ee
The latter identity is due to the independence of the random variables $Y_1,Y_2,\dots$
The sum over $k$ is a geometric series, and one obtains a useful expression:
\be
\bbP_{{\rm RA}(\nu_1)}(T,U \in \text{ same partition element}) = \Bigl( 2 \frac{\bbE_{\nu_1} Y}{\bbE_{\nu_1} Y^2} - 1 \Bigr)^{-1}.
\ee

The case that is relevant for our purpose is when $\nu_1$ is a Beta($\vartheta$) random variable. That is, the random number $Y$ has distribution Beta($\vartheta$) if
\be
\bbP_{\nu_1}(Y > s) = (1-s)^\vartheta,
\ee
for $0 \leq s \leq 1$. Its probability density function is $\vartheta (1-s)^{\vartheta-1}$, so that
\be
\bbE_{\nu_1}(f(Y)) = \int_0^1 f(s) \vartheta (1-s)^{\vartheta-1} \dd s.
\ee
The residual allocation model where $\nu_1$ is the measure of a Beta($\vartheta$) random variable, is called the {\bf Griffiths-Engen-McCloskey} GEM($\vartheta$) distribution. It appears in mathematical biology. Rearranging the unordered partition $(X_1,X_2,\dots)$ in decreasing order, we get a random partition with {\bf Poisson-Dirichlet} PD($\vartheta$) distribution.

\subsection{Kingman's representation of Poisson-Dirichlet}
\label{sec Kingman}

We now discuss another expression of the Poisson-Dirichlet distribution that is due to Kingman \cite{Kin}. It is useful in order to calculate moments.

Let $Z_{1}, \dots, Z_k$ be i.i.d.\ random variables with Gamma$(\frac\vartheta k,1)$ distribution (that is, their probability density function is $s^{\frac\vartheta k - 1} \e{-s} / \Gamma(\frac\vartheta k)$ for $0 \leq s < \infty$). Let $S = Z_{1} + \dots + Z_k$. Consider the sequence
\be
\Bigl( \frac{Z_{1}}S, \dots, \frac{Z_k}S \Bigr)
\ee
and reorder it in decreasing order, so it forms a random partition of $[0,1]$. As $k\to\infty$, this partition turns out to converge to PD$(\vartheta)$. The following two observations are keys to our calculations:
\begin{itemize}
\item $S$ is a Gamma$(\vartheta,1)$ random variable;
\item $S$ is independent of $(\frac{Z_{1}}S, \dots, \frac{Z_k}S)$.
\end{itemize}
The first observation is easy to verify. As for the second observation, we have for arbitrary functions $f : \bbR \to \bbR$ and $g : \bbR^k \to \bbR$,
\be
\begin{split}
\bbE_{\{Z_i\}_{i=1}^k} \Bigl( f(S) \, & g\bigl( \tfrac{Z_1}S,\dots,\tfrac{Z_k}S \bigr) \Bigr) = \int_0^\infty \dd z_1 \dots \dd z_k \prod_{i=1}^k \frac{z_i^{\frac\vartheta k - 1} \e{-z_i}}{\Gamma(\frac\vartheta k)} f \Bigl( \sum z_i \Bigr) g \bigl( \tfrac{z_1}{\sum z_i}, \dots, \tfrac{z_k}{\sum z_i} \bigr) \\
&= \int_0^\infty \dd s \int_0^\infty \dd z_1 \dots \dd z_k \prod_{i=1}^k \frac{z_i^{\frac\vartheta k - 1} \e{-z_i}}{\Gamma(\frac\vartheta k)} f(s) g \bigl( \tfrac{z_1}s, \dots, \tfrac{z_k}s \bigr) \, \delta \Bigl( s-\sum z_i \Bigr) \\
&= \int_0^\infty \dd s \, s^\vartheta \e{-s} f(s) \int_0^\infty \dd y_1 \dots \dd y_k \prod_{i=1}^k \frac{y_i^{\frac\vartheta k - 1}}{\Gamma(\frac\vartheta k)} g(y_1, \dots, y_k) \, \delta \Bigl( s \bigl( 1-\sum y_i \bigr) \Bigr).
\end{split}
\ee
We made the change of variables $y_i = z_i / s$. We now use $\delta(sx) = \frac1s \delta(x)$, which can be seen using such representation of the Dirac function as $\delta(x) = \lim_{n\to\infty} \sqrt{\frac n\pi} \e{-n x^2}$. We get
\bm
\bbE_{\{Z_i\}_{i=1}^k} \Bigl( f(S) \, g\bigl( \tfrac{Z_1}S,\dots,\tfrac{Z_k}S \bigr) \Bigr) = \int_0^\infty \dd s \frac{s^{\vartheta-1} \e{-s}}{\Gamma(\vartheta)} f(s) \\
\cdot \Gamma(\vartheta) \int_0^\infty \dd y_1 \dots \dd y_k \prod_{i=1}^k \frac{y_i^{\frac\vartheta k - 1}}{\Gamma(\frac\vartheta k)} g(y_1, \dots, y_k) \, \delta \Bigl( 1-\sum y_i \Bigr).
\end{multline}
The first line of the right side is equal to $\bbE_{{\rm Gamma}(\vartheta,1)}(f)$. The second line of the right side does not depend on $s$; by looking at the special case $f\equiv1$, this must be equal to the expectation of the function $g$.

We check in Section \ref{sec random permutations} that the ordered sequence has Poisson-Dirichlet distribution with parameter $\vartheta$ in the limit $k\to\infty$.

\subsection{Moments of the Poisson-Dirichlet distribution}

For given integers $n_{1}, \dots, n_{\ell} \geq 0$, using the independence of $S$ from the partition, we have
\be
\begin{split}
\bbE_{{\rm PD}(\vartheta)} \Bigl( \sumtwo{j_{1}, \dots, j_{\ell} \geq 1}{\rm distinct} X_{j_{1}}^{n_{1}} \dots X_{j_{\ell}}^{n_{\ell}} \Bigr) &= \lim_{k\to\infty} \frac{k!}{(k-\ell)!} \; \bbE_{\{Z_i\}_{i=1}^k} \Bigl( \Bigl( \frac{Z_{1}}S \Bigr)^{n_{1}} \dots \Bigl( \frac{Z_{\ell}}S \Bigr)^{n_{\ell}} \Bigr) \\
&= \lim_{k\to\infty} \frac{k!}{(k-\ell)!} \; \frac{\bbE_{\{Z_i\}_{i=1}^k} \bigl( S^{n_{1} + \dots + n_{\ell}} (\frac{Z_{1}}S)^{n_{1}} \dots (\frac{Z_{\ell}}S)^{n_{\ell}} \bigr)}{\bbE_{\{Z_i\}_{i=1}^k} ( S^{n_{1} + \dots + n_{\ell}})} \\
&= \lim_{k\to\infty} \frac{k!}{(k-\ell)!} \; \frac{\Gamma(\vartheta) \, \bbE_{\{Z_i\}_{i=1}^k} \bigl( Z_{1}^{n_{1}} \dots Z_{\ell}^{n_{\ell}} \bigr)}{\Gamma(\vartheta + n_{1}+\dots+n_{\ell})}.
\end{split}
\ee
We also used $\bbE_{\{Z_i\}_{i=1}^k} (S^{a}) = \Gamma(\vartheta+a) / \Gamma(\vartheta)$. Since the $Z_{i}$s are independent,
\be
\bbE_{\{Z_i\}_{i=1}^k} \bigl( Z_{1}^{n_{1}} \dots Z_{\ell}^{n_{\ell}} \bigr) = \prod_{i=1}^{\ell} \frac{\Gamma(\vartheta/k + n_{i})}{\Gamma(\vartheta/k)}.
\ee
Recall that $\Gamma(\vartheta/k) \sim k/\vartheta$ as $k\to\infty$, so that $\frac{k!}{(k-\ell)! \Gamma(\vartheta/k)^\ell} \to \vartheta^{\ell}$. We obtain
\be
\label{c'est important !}
\bbE_{{\rm PD}(\vartheta)} \Bigl( \sumtwo{j_{1}, \dots, j_\ell \geq 1}{\rm distinct} X_{j_{1}}^{n_{1}} \dots X_{j_\ell}^{n_\ell} \Bigr) = \frac{\vartheta^\ell \, \Gamma(\vartheta) \, \Gamma(n_{1}) \dots \Gamma(n_\ell)}{\Gamma(\vartheta + n_{1} + \dots + n_\ell)}.
\ee
This important formula appears in \cite{NCSOS}. Its derivation there is different; it involves another loop soup model, assumes the presence of Poisson-Dirichlet, and uses a ``supersymmetry'' method.


\subsection{Expectation of functions of partition elements}

We now consider the Poisson-Dirichlet expectation of a general smooth function $f$ that satisfies $f(0)=1$.
Let $(a_k)_{k\geq1}$ be Taylor coefficients such that the following function has radius of convergence greater than 1:
\be
f(s) = 1 + \sum_{k\geq1} a_k s^k.
\ee
Then, using \eqref{c'est important !},
\be
\label{ca aussi !}
\begin{split}
\bbE_{{\rm PD}(\vartheta)} \Bigl( \prod_{i\geq1} f(X_i) \Bigr) &= \sum_{n\geq0} \frac1{n!} \sumtwo{i_1, \dots, i_n \geq 1}{\rm distinct} \sum_{k_1,\dots,k_n\geq1} a_{k_1} \dots a_{k_n} \bbE_{{\rm PD}(\vartheta)} \Bigl( X_{i_1}^{k_1} \dots X_{i_n}^{k_n} \Bigr) \\
&= \sum_{n\geq0} \frac1{n!}  \sum_{k_1,\dots,k_n\geq1} a_{k_1} \dots a_{k_n} \frac{\vartheta^{n} \, \Gamma(\vartheta) \, \Gamma(k_{1}) \dots \Gamma(k_{n})}{\Gamma(\vartheta + k_{1} + \dots + k_{n})}
\end{split}
\ee

Let us apply this formula to a special case that will be useful in Section \ref{sec conseq}, namely $f(s) = \cosh(bs)$ with $b$ a parameter. The Taylor coefficients are
\be
a_k = \begin{cases} \frac{b^k}{k!} & \text{ if $k$ is even,} \\ 0 & \text{ if $k$ is odd.} \end{cases}
\ee
Then
\be
\label{calc cosh}
\begin{split}
\bbE_{{\rm PD}(\vartheta)} \Bigl( \prod_{i\geq1} \cosh(b X_i) \Bigr) &= 1 + \sum_{n\geq1} \frac{\vartheta^n}{n!} \sumtwo{k_1,\dots,k_n\geq2}{{\rm even}} \frac1{k_1 \dots k_n} \frac{\Gamma(\vartheta)}{\Gamma(\vartheta + \sum_i k_i)} b^{\sum_i k_i} \\
&= 1 + \sum_{n\geq1} \frac{\vartheta^n}{n!} \sum_{r\geq n} \sumtwo{\ell_1,\dots,\ell_n\geq1}{\ell_1+\dots+\ell_n=r} \frac1{2^n \ell_1 \dots \ell_n} \frac{\Gamma(\vartheta)}{\Gamma(\vartheta + 2r)} b^{2r} \\
&= 1 + \sum_{r\geq1} \frac{\Gamma(\vartheta) b^{2r}}{\Gamma(\vartheta + 2r)} \sum_{n=1}^r \frac{(\vartheta/2)^n}{n!} \sumtwo{\ell_1,\dots,\ell_n\geq1}{\ell_1+\dots+\ell_n=r} \frac1{\ell_1 \dots \ell_n} \\
&= \frac{\Gamma(\vartheta)}{\Gamma(\vartheta/2)} \sum_{r\geq0} \frac{\Gamma(\vartheta/2 + r)}{r! \Gamma(\vartheta+2r)} b^{2r}.
\end{split}
\ee
We used the identity
\be
\label{cool identity}
\sum_{n=1}^r \frac{\theta^n}{n!} \sumtwo{\ell_1,\dots,\ell_n\geq1}{\ell_1+\dots+\ell_n=r} \frac1{\ell_1 \dots \ell_n} = \frac{\Gamma(\theta+r)}{r! \Gamma(\theta)}.
\ee


\section{Random permutations}
\label{sec random permutations}

Random permutations provide a convenient mean to understanding random partitions, their distributions, and the split-merge process. We should point out that, in this section and the next one, there is no space --- we are dealing with mean-field models. This is nonetheless directly relevant to spatial systems in dimensions three or larger, as is explained in Section \ref{sec heuristics}.

\subsection{The Ewens distribution and natural extensions}
We consider four ensembles of random permutations, with fixed or variable number of elements and number of cycles. Let $\caS_n^k$ denote the set of permutations of $n$ elements and $k$ cycles, and let
\be
\caS_n = \bigcup_{k=1}^n \caS_n^k, \qquad \caS^k = \bigcup_{n\geq1} \caS_n^k, \qquad \caS = \bigcup_{n\geq1} \bigcup_{k=1}^n \caS_n^k.
\ee
Given a permutation $\sigma \in \caS$, we let $N(\sigma)$ and $K(\sigma)$ denote its number of elements and its number of cycles, respectively. It is worth recalling that the number of permutations with $n$ elements and $k$ (labelled) cycles of lengths $m_1,\dots,m_k$ is equal to
\be
\label{number permutations}
\left( \begin{matrix} n \\ m_1 \dots m_k \end{matrix} \right) \prod_{i=1}^k (m_i-1)! = \frac{n!}{m_1 \cdots m_k}.
\ee

The sets $\caS_n^k$, $\caS_n$, $\caS^k$, and $\caS$ are reminiscent of the microcanonical, canonical, and grand-canonical ensembles of particle systems in statistical physics, with number of elements and cycles playing a somewhat similar r\^ole as energy and number of particles.
We consider probability distributions on these sets, namely
\begin{align}
&\bbP_{n,k}(\sigma) = \frac1{n! \, Z_{n,k}} &\text{for } \sigma \in \caS_n^k; \label{distr n k} \\
&\bbP_{n,\theta}(\sigma) = \frac1{n! \, Z_{n,\theta}} \theta^{K(\sigma)} &\text{for } \sigma \in \caS_n; \label{distr n theta} \\
&\bbP_{z,k}(\sigma) = \frac1{Z_{z,k}} \, \frac{z^{N(\sigma)}}{N(\sigma)!} &\text{for } \sigma \in \caS^k; \label{distr z k} \\
&\bbP_{z,\theta}(\sigma) = \frac1{Z_{z,\theta}} \, \frac{z^{N(\sigma)} \theta^{K(\sigma)}}{N(\sigma)!} &\text{for } \sigma \in \caS. \label{distr z theta}
\end{align}
The second distribution, $\bbP_{n,\theta}$, is the Ewens distribution that initially appeared in mathematical biology.
These distributions are related as follows:
\begin{align}
&\bbP_{n,\theta}(\cdot | K(\sigma)=k) = \bbP_{n,k}(\cdot); \\
&\bbP_{z,k}(\cdot | N(\sigma)=n) = \bbP_{n,k}(\cdot); \\
&\bbP_{z,\theta}(\cdot | N(\sigma)=n) = \bbP_{n,\theta}(\cdot); \\
&\bbP_{z,\theta}(\cdot | K(\sigma)=k) = \bbP_{z,k}(\cdot); \\
&\bbP_{z,\theta}(\cdot | N(\sigma)=n, K(\sigma)=k) = \bbP_{n,k}(\cdot).
\end{align}

The last three normalisations can be calculated explicitly. Using Eq.\ \eqref{number permutations}, we have
\be
\label{calc Z}
\begin{split}
Z_{z,\theta} &= \sum_{n\geq1} \frac{z^n}{n!} \sum_{\sigma\in\caS_n} \theta^{K(\sigma)} \\
&= \sum_{n\geq1} \frac{z^n}{n!} \sum_{k\geq1} \frac{\theta^k}{k!} \sumtwo{m_1,\dots,m_k\geq1}{m_1+\dots+m_k=n} \frac{n!}{m_1 \dots m_k} \\
&= \sum_{k\geq1} \frac{\theta^k}{k!} \Bigl( \sum_{m\geq1} \frac{z^m}m \Bigr)^k \\
&= \exp \bigl( -\theta \log(1-z) \bigr) - 1\\
&= (1-z)^{-\theta} - 1.
\end{split}
\ee
We have the relations
\be
Z_{z,\theta} = \sum_{n\geq1} z^n Z_{n,\theta} = \sum_{k\geq1} \theta^k Z_{z,k},
\ee
so we get $Z_{n,\theta}$ by differentiating $n$ times with respect to $z$, and we get $Z_{z,k}$ by looking at the $k$th coefficient in the middle line of Eq.\ \eqref{calc Z}; explicitly,
\begin{align}
&Z_{n,\theta} = \frac{\theta (\theta+1) \dots (\theta+n-1)}{n!} = \frac{n^{\theta-1}}{\Gamma(\theta)} \bigl( 1 + o(1) \bigr), \label{Zntheta} \\
&Z_{z,k} = \frac1{k!} \bigl( -\log(1-z) \bigr)^k.
\end{align}
The first normalisation, $Z_{n,k}$ does not have an explicit expression. The numbers $n! Z_{n,k}$ are known as Stirling numbers of the first kind. The following asymptotic behaviour is useful for our purpose; if $k = \lambda \log n$, we have \cite{Hwa}
\be
\label{asymptotic Stirling}
Z_{n,k} = \frac1{\Gamma(1+\lambda)} \, \frac{(\log n)^{k-1}}{n \, (k-1)!} \bigl( 1 + o(1) \bigr).
\ee

The relevant limits are
\begin{itemize}
\item $n,k\to\infty$ with $k = \theta \log n$ for some fixed parameter $\theta$;
\item $n\to\infty$ with fixed $\theta$;
\item $k \to \infty$ and $z\to1-$ with $z = \e{-\e{-k/\theta}}$ for some fixed parameter $\theta$;
\item $z \to 1-$ with fixed $\theta$.
\end{itemize}

We now check that, as $z\to1-$ with $\bbP_{z,\theta}$, the number of elements diverges like $(-\log z)^{-1}$ and, in this scaling, behaves like a Gamma random variable; see Eq.\ \eqref{N is Gamma} below. First,
\be
\bbP_{z,\theta}(N(\sigma) \leq \tfrac a{-\log z}) = \frac1{Z_{z,\theta}} \sum_{n=1}^{a/-\log z} z^n Z_{n,\theta}.
\ee
As $z\to1-$, only large $n$ contribute to the sum and we can use the asymptotics in \eqref{Zntheta}. We get
\be
\begin{split}
\lim_{z\to1-} \bbP_{z,\theta}(N(\sigma) \leq \tfrac a{-\log z}) &= \lim_{z\to1-} \frac{(1-z)^\theta}{\Gamma(\theta)}  \sum_{n=1}^{a/-\log z} n^{\theta-1} z^n \\
&= \lim_{z\to1-} \frac{(1-z)^\theta}{\Gamma(\theta)} \int_0^{a/-\log z} s^{\theta-1} \e{-(\log z) s} \dd s \\
&= \lim_{z\to1-} \Bigl( \frac{1-z}{-\log z} \Bigr)^\theta \frac1{\Gamma(\theta)} \int_0^a s^{\theta-1} \e{-s} \dd s.
\end{split}
\ee
We obtain that $N(\sigma)$ is a Gamma random variable multiplied by $(-\log z)^{-1}$; namely, we have for all $a>0$ that
\be
\label{N is Gamma}
\lim_{z\to1-} \bbP_{z,\theta}(N(\sigma) \leq \tfrac a{-\log z}) = \frac1{\Gamma(\theta)} \int_0^a s^{\theta-1} \e{-s} \dd s.
\ee

A similar statement holds with $\bbP_{z,k}$ with suitable limits $z\to1-$ and $k\to\infty$. Let $z(k) = \e{-\e{-k/\theta}}$.
\be
\begin{split}
\bbP_{z(k),k}(N(\sigma) \leq a \e{k/\theta}) &= \sum_{n=1}^{a \e{k/\theta}} \frac{z(k)^n Z_{n,k}}{Z_{z(k),k}} \\
&= \bigl( 1 + o(1) \bigr) \sum_{n=1}^{a \e{k/\theta}} \e{-n \e{-k/\theta}} \frac{(\log n)^{k-1}}{n \, (k-1)!} \, \frac1{\Gamma(\theta+1)} \, \frac{k!}{(k/\theta)^k}.
\end{split}
\ee
We used the asymptotic result \eqref{asymptotic Stirling} and also
\be
\label{Zzkk}
Z_{z(k),k} = \frac1{k!} \Bigl( \frac k\theta \Bigr)^k \bigl( 1 + O(\e{-k/\theta}) \bigr).
\ee
This can be justified by first showing that $K(\sigma)/\theta \log N(\sigma)$ tends to 1 with probability 1; this is not too difficult, but we do not write it down here. Then
\be
\begin{split}
\bbP_{z(k),k}(N(\sigma) \leq a \e{k/\theta}) &= \bigl( 1 + o(1) \bigr) \frac\theta{\Gamma(\theta+1)} \sum_{n=1}^{a \e{k/\theta}} \e{-n \e{-k/\theta}} \Bigl( \frac{\theta \log n}k \Bigr)^{k-1} \frac1n \\
&= \bigl( 1 + o(1) \bigr) \frac1{\Gamma(\theta)} \int_0^{a \e{k/\theta}} \Bigl( \frac{\theta \log n}k \Bigr)^{k-1} \e{-n \e{-k/\theta}} \dd n \\
&= \bigl( 1 + o(1) \bigr) \frac1{\Gamma(\theta)} \int_0^a \frac1s \underbrace{\Bigl(1 + \frac{\theta \log s}k \Bigr)^{k-1}}_{\to s^\theta} \e{-s} \dd s.
\end{split}
\ee
We obtain that $N$ behaves like a Gamma($\theta,1$) random variable multiplied by $\e{k/\theta}$: For $a>0$,
\be
\lim_{k\to\infty} \bbP_{z(k),k}(N(\sigma) \leq a \e{k/\theta}) =  \frac1{\Gamma(\theta)} \int_0^a s^{\theta-1} \e{-s} \dd s.
\ee

We now verify that the distribution of cycle lengths is asymptotically equivalent to i.i.d.\ Gamma($\frac\theta k,\e{-k/\theta}$) random variables. Together with the result of the next subsection, this justifies Kingman's representation of Poisson-Dirichlet described in Section \ref{sec Kingman}.

The probability to obtain a permutation with $k$ cycles of lengths $m_1,\dots,m_k$ is, with $n = \sum_{i=1}^k m_i$,
\be
\label{prob mms}
\begin{split}
\bbP_{z(k),k}(m_1,\dots,m_k) &= \bigl(1+o(1)\bigr) \frac{k!}{(k\theta)^k} \frac{z(k)^n}{n!} \frac{n!}{k! m_1 \dots m_k} \\
&= \bigl(1+o(1)\bigr) \frac1{(k/\theta)^k} \prod_{i=1}^k \frac{\e{-m_i \e{-k/\theta}}}{m_i}.
\end{split}
\ee
We used Eqs \eqref{number permutations} and \eqref{Zzkk}.

On the other hand, the probability that $k$ i.i.d.\ Gamma($\frac\theta k,\e{-k/\theta}$) random variables take values in $[m_1,m_1+1], \dots, [m_k,m_k+1]$ is equal to
\be
\label{ca semble different}
\bigl(1+o(1)\bigr) \prod_{i=1}^k \frac{m_i^{\frac\theta k-1}}{\Gamma(\theta/k)} \e{-m_i \e{-k/\theta}}.
\ee
In order to match this with \eqref{prob mms}, observe that
\be
\label{LLN}
\begin{split}
\log \prod_{i=1}^k m_i^{\frac\theta k} &= \tfrac\theta k \sum_{i=1}^k \log m_i \\
&\approx \theta \bbE_{{\rm Gamma}(\frac\theta k,\e{-k/\theta})} \log X \\
&= \theta \psi(\tfrac\theta k) + k,
\end{split}
\ee
where $\psi(\cdot) = \Gamma'(\cdot) / \Gamma(\cdot)$ is the digamma function. For large $k$, we have the asymptotics
\be
\label{asymptotics gamma}
\begin{split}
&\Gamma(\tfrac\theta k)^k = \bigl( \tfrac k\theta \bigr)^k \e{-\theta\gamma} \bigl(1+o(1)\bigr), \\
&\psi(\tfrac\theta k) = -\tfrac k\theta - \gamma +o(1).
\end{split}
\ee
Here, $\gamma$ is Euler-Mascheroni constant. Using \eqref{LLN} and \eqref{asymptotics gamma} in \eqref{ca semble different}, we get Eq.\ \eqref{prob mms}. This shows that the random partition from $(\frac{Z_1}S,\dots,\frac{Z_k}S)$ has asymptotically the same distribution as the one from the cycle lengths of a random permutation distributed according to $\bbP_{z(k),k}$. There remains to check that the latter has Poisson-Dirichlet distribution.

\subsection{Cycle structure of Ewens permutations}

Given $\sigma \in \caS$, let $L_1(\sigma)$ be the length of the cycle that contains the element 1; $L_2(\sigma)$ the length of the cycle that contains the smallest element that is not in the first cycle; $L_3(\sigma)$ the length of the cycle that contains the smallest element that is not in the first two cycles; etc... Then $\sum_{i=1}^{K(\sigma)} L_i(\sigma) = N(\sigma)$ for all $\sigma \in \caS$, and $(\frac{L_1}{N(\sigma)}, \dots, \frac{L_{K(\sigma)}}{N(\sigma)})$ is an unordered partition of $[0,1]$. It turns out that, if $\sigma$ is chosen randomly according to the measures \eqref{distr n k}--\eqref{distr z theta}, and taking appropriate limits, the distribution of cycle lengths converges to GEM. This  is well-known in the case of the Ewens measure \eqref{distr n theta}, see \cite{ABT}, and we show it here for the other distributions.

We start with the distribution with fixed $n,k$ given in \eqref{distr n k}; we take $k = \theta \log n$ and consider the limit $n\to\infty$. The first step is to show that $L_1/n$ converges to a Beta random variable with parameter $\theta$. We have
\be
\begin{split}
\bbP_{n,k} \Bigl( \frac{L_1(\sigma)}n \leq a \Bigr) &= \sum_{j=1}^{an} \left( \begin{matrix} n-1 \\ j-1 \end{matrix} \right) (j-1)! \frac{Z_{n-j,k-1}}{Z_{n,k}} (n-j)! \\
&= \frac1n  \sum_{j=1}^{an} \frac{Z_{n-j,k-1}}{Z_{n,k}} \\
&= \frac1n \sum_{j=1}^{an} \frac{(\log(n-j))^{k-2}}{(n-j) (k-2)!} \frac1{\Gamma(1+\theta)} \frac{n (k-1)!}{(\log n)^{k-1}} \Gamma(1+\theta) \bigl( 1 + o(1) \bigr) \\
&= \sum_{j=1}^{an} \frac{k-1}{n-j} \, \frac1{\log n} \Bigl( \frac{\log n + \log(1 - \frac jn)}{\log n} \Bigr)^{k-2} \bigl( 1 + o(1) \bigr).
\end{split}
\ee
We used the asymptotic result \eqref{asymptotic Stirling}. We have $\frac{k-1}{\log n} = \theta (1+o(1))$ and
\be
\Bigl( \frac{\log n + \log(1 - \frac jn)}{\log n} \Bigr)^{k-2} = \e{\theta \log (1-\frac jn)} \bigl( 1 + o(1) \bigr)  = \bigl( 1-\tfrac jn \bigr)^\theta \bigl( 1 + o(1) \bigr).
\ee
We get
\be
\label{limit first cycle}
\begin{split}
\bbP_{n,k} \Bigl( \frac{L_1(\sigma)}n \leq a \Bigr) &= \frac\theta n \sum_{j=1}^{an} \bigl( 1-\tfrac jn \bigr)^{\theta-1} \bigl( 1 + o(1) \bigr) \\
&\substack{n\to\infty \\ \longrightarrow} \theta \int_0^a (1-s)^{\theta-1} \dd s.
\end{split}
\ee
The latter expression is indeed equal to $\bbP_{{\rm Beta}(\theta)}(X \leq a)$. Next, we consider the joint distribution of the lengths of the first $j$ cycles; we keep $j$ fixed and take the limit $n\to\infty$. We have
\be
\begin{split}
&\bbP_{n,k} \Bigl( \frac{L_1(\sigma)}n \leq a_1, \dots, \frac{L_j(\sigma)}{n-L_1-\dots-L_{j-1}} \leq a_j \Bigr) \\
&= \sum_{m_1=1}^{a_1 n} \dots \sum_{m_j=1}^{a_j (n-m_1-\dots-m_{j-1})} \bbP_{n,k}(L_1=m_1,\dots,L_j=m_j) \\
&= \sum_{m_1=1}^{a_1 n} \dots \sum_{m_j=1}^{a_j (n-m_1-\dots-m_{j-1})} \bbP_{n,k}(L_1=m_1,\dots,L_{j-1}=m_{j-1}) \\
&\hspace{4cm} \cdot \bbP_{n,k}(L_j=m_j | L_1=m_1,\dots,L_{j-1}=m_{j-1}).
\end{split}
\ee
We now use self-similarity for the last term; having determined the lengths of the first $j-1$ cycles, the distribution of the length of the $j$th cycles is the same but with less elements:
\be
\bbP_{n,k}(L_j \leq a_j (n-m_1-\dots-m_{j-1} | L_1=m_1,\dots,L_{j-1}=m_{j-1}) = \bbP_{n',k'}(L_j \leq a_j n'),
\ee
with $n' = n-m_1-\dots-m_{j-1}$ and $k'=k-j+1$. Since $j$ is fixed, the limit $k = \theta \log n \to \infty$ corresponds to $k' = \theta \log n' \to \infty$; using the above result \eqref{limit first cycle}, we have
\be
\limtwo{k,n\to\infty}{k=\theta\log n} \bbP_{n',k'}(L_j \leq a_j n') = \bbP_{{\rm Beta}(\theta)}(X \leq a_j).
\ee
This allows to prove by induction that
\be
\limtwo{k,n\to\infty}{k=\theta\log n} \bbP_{n,k} \Bigl( \frac{L_1(\sigma)}n \leq a_1, \dots, \frac{L_j(\sigma)}{n-L_1-\dots-L_{j-1}} \leq a_j \Bigr) = \prod_{i=1}^j \bbP_{{\rm Beta}(\theta)}(X \leq a_i).
\ee
This means that the joint distribution of $(\frac{L_1}n, \frac{L_2}n, \dots)$ is GEM($\theta$).

As pointed out before, the same result holds with the distribution $\bbP_{n,\theta}$ on $\caS_n$. 
This can be extended to the measure $\bbP_{z,\theta}$ on $\caS$ in the limit $z=\to1-$. Indeed, we have
\be
\begin{split}
\bbP_{z,\theta} &\Bigl( \frac{L_1}{N} \leq a_1, \dots, \frac{L_j}{N-L_1-\dots-L_{j-1}} \leq a_j \Bigr) \\
&= \sum_{n\geq1} \bbP_{z,\theta}(N=n) \, \bbP_{n,\theta} \Bigl( \frac{L_1}{n} \leq a_1, \dots, \frac{L_j}{n-L_1-\dots-L_{j-1}} \leq a_j \Bigr).
\end{split}
\ee
We have seen that $N(\sigma)$ diverges as $z\to-1$ (as $(-\log z)^{-1}$), so only large $n$ matter, for which the conditional probability approaches the product of Beta probabilities.

\section{Split-merge process}
\label{sec split-merge}

The split-merge process, also called {\it coagulation-fragmentation}, is a discrete-time stochastic process on the set of partitions of the interval $[0,1]$. It involves two parameters $g_{\rm s}, g_{\rm m} \in [0,1]$. Given a partition $(\lambda_1,\lambda_2,\dots)$ at time $t\in\bbN$, the partition at time $t+1$ is obtained as follows. Choose two numbers in $[0,1]$, uniformly at random. Then
\begin{itemize}
\item if they fall in the same partition element, we split this element with probability $g_{\rm s}$, uniformly;
\item if they fall in distinct partition elements, we merge these elements with probability $g_{\rm m}$.
\end{itemize}
After rearranging in decreasing order, we get the partition for time $t+1$. This process is illustrated in Fig.\ \ref{fig split merge}.

\bfig
\includegraphics[height=50mm]{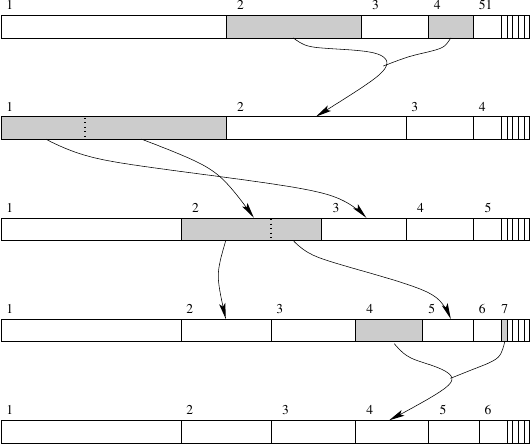}
\caption{Illustration for the split-merge process. This sequence involves a merge; a split; another split; another merge.}
\label{fig split merge}
\efig

There is a continuous-time equivalent process, where an element $\lambda_j$ splits at rate $\lambda_j^2 g_{\rm s}$; and elements $\lambda_i, \lambda_j$ (with $i \neq j$) merge at rate $2\lambda_i \lambda_j g_{\rm m}$. This means that if $(\lambda_1, \lambda_2, \dots)$ is the partition at time $t \in [0,\infty)$, then during the tiny interval $[t,t+\dd t]$,
\begin{itemize}
\item $\lambda_j$ splits with probability $\lambda_j^2 g_{\rm s} \dd t$;
\item $\lambda_i, \lambda_j$ (with $i \neq j$) merge with probability $2\lambda_i \lambda_j g_{\rm m} \dd t$;
\item no changes occur with probability $1 - \sum_{j\geq1} \lambda_j^2 g_{\rm s} \dd t - \sum_{i<j} 2\lambda_i \lambda_j g_{\rm m} \dd t$.
\end{itemize}

We now check that the invariant measure of the split-merge process is Poisson-Dirichlet with parameter $\vartheta = g_{\rm s} / g_{\rm m}$. We first give an indirect proof using a process on permutations; the invariant measure is Ewens; when projected onto partitions, in the limit of infinitely-many elements, we get the split-merge process and the GEM or PD distributions. The second proof is more direct but it is more cumbersome and we only discuss it in the case $g_{\rm s} = g_{\rm m} = 1$. Relevant references for this section include \cite{Ber, Tsi, Sch, DMZZ}.

\subsection{Markov process on $\caS_n$}

Let $\tau_{ij}$ denote the transposition of elements $i,j \in \{1,\dots,n\}$. Recall that $K(\sigma)$ is the number of cycles of the permutation $\sigma$. One easily checks that, if $i,j$ belong to distinct cycles of $\sigma$, then $i,j$ belong to the same cycle of $\tau_{ij} \circ \sigma$; conversely, if $i,j$ belong to the same cycle of $\sigma$, then $i,j$ belong to distinct cycles of $\tau_{ij} \circ \sigma$. We always have $K(\tau_{ij} \circ \sigma) = K(\sigma) \pm 1$.

The process we consider is a simple process that involves products of transpositions. Let $\sigma_t$ denote the permutation at time $t$. Choose $i,j \in \{1,\dots,n\}$ at random, with $i \neq j$.
\begin{itemize}
\item If $\tau_{ij}$ spits a cycle, i.e.\ $K(\tau_{ij} \circ \sigma_t) = K(\sigma_t) + 1$, then $\sigma_{t+1} = \tau_{ij} \circ \sigma$ with probability $g_{\rm s}$; $\sigma_{t+1} = \sigma$ otherwise.
\item If $\tau_{ij}$ merges two cycles, i.e.\ $K(\tau_{ij} \circ \sigma_t) = K(\sigma_t) - 1$, then $\sigma_{t+1} = \tau_{ij} \circ \sigma$ with probability $g_{\rm m}$; $\sigma_{t+1} = \sigma$ otherwise.
\end{itemize}
The transition matrix is
\be
T(\sigma; \tau_{ij} \circ \sigma) = \frac1{\frac12 n (n-1)} \begin{cases} g_{\rm s} & \text{if } K(\tau_{ij} \circ \sigma_t) = K(\sigma_t) + 1, \\ g_{\rm m} & \text{if } K(\tau_{ij} \circ \sigma_t) = K(\sigma_t) - 1, \end{cases}
\ee
and $T(\sigma;\sigma) = 1 - \sum_{i<j} T(\sigma; \tau_{ij} \circ \sigma)$. Let $p_t(\sigma)$ denote the probability of the permutation $\sigma$ at time $t$; the probability at time $t+1$ satisfies
\be
\label{prob t+1}
p_{t+1}(\sigma) = p_t(\sigma) T(\sigma;\sigma) + \sum_{i<j} p_t(\tau_{ij} \circ \sigma) T(\tau_{ij} \circ \sigma; \sigma).
\ee
Indeed, $\tau_{ij} \circ \sigma$ is the permutation that gives $\sigma$ if we apply $\tau_{ij}$. The measure $p_t$ is invariant if $p_{t+1}=p_t$. A sufficient condition for this is that it satisfies the {\bf detailed balance condition}
\be
\label{detailed balance}
p(\sigma) T(\sigma; \tau_{ij} \circ \sigma) = p(\tau_{ij} \circ \sigma) T(\tau_{ij} \circ \sigma; \sigma).
\ee
Indeed, inserting this identity in \eqref{prob t+1} yields $p_{t+1}=p_t$.

One easily checks that the Ewens measure $\bbP_{n,\theta} = \frac1{Z_{n,\theta}} \theta^{K(\sigma)}$ satisfies the detailed balance condition: Assume that $K(\tau_{ij} \circ \sigma) = K(\sigma)+1$; then
\be
\bbP_{n,\theta}(\tau_{ij} \circ \sigma) T(\tau_{ij} \circ \sigma; \sigma) = \theta \bbP_{n,\theta}(\sigma) \frac1{\frac12 n (n-1)} g_{\rm m} = \theta \frac{g_{\rm m}}{g_{\rm s}} \bbP_{n,\theta}(\sigma) T(\sigma; \tau_{ij} \circ \sigma).
\ee
This is identical to \eqref{detailed balance} provided $\boxed{\theta = \frac{g_{\rm s}}{g_{\rm m}}}$. The same argument applies to the case $K(\tau_{ij} \circ \sigma) = K(\sigma)-1$.

Permutations of $\caS_n$ can be projected onto set partitions on $\{1,\dots,n\}$, with sets given by permutation cycles. The Markov process above gives a Markov process on set partitions: Choose $i,j \in \{1,\dots,n\}$, $i\neq j$; if they fall in the same set, we split it with probability $g_{\rm s}$; if they fall in distinct sets, we merge them with probability $g_{\rm m}$.

Further, set partitions can projected onto integer partitions, according to the cardinalities of the sets. The Markov process gives a split-merge process that is still Markov and is a discretised version of the one described above. Dividing the elements by $n$, and letting $n\to\infty$, we recover the standard split-merge process.

As $n\to\infty$, the cycle lengths of Ewens random permutations with parameter $\theta$ have Poisson-Dirichlet distribution with the same parameter, $\vartheta=\theta$. Since cycle lengths satisfy a split-merge process, we can conclude that its invariant measure is Poisson-Dirichlet with parameter $\vartheta = \frac{g_{\rm s}}{g_{\rm m}}$.

All this is well-known in mathematical biology and probability theory. We refer to \cite{Tsi,Pit,DMZZ,Ber} for further information, including mathematical results about the delicate issue of uniqueness of the invariant measure.

\subsection{Split-merge process for GEM}

We now consider unordered partitions and introduce a modified split-merge process whose invariant measure is GEM($\vartheta$). If we project onto ordered partitions, we recover the usual split-merge process. Since GEM projects onto PD, this indeed proves that PD is invariant for split-merge. This proof could perhaps be extended to the case $g_{\rm s}, g_{\rm m} \neq 1$, but this remains to be clarified.

It is convenient to work with integer partitions, so we deal with probabilities rather than densities, and we avoid the tiny but numerous elements at the accumulation point. Let $n \in \bbN$ be a large number and let $\caM_n$ denote the set of unordered integer partitions of $n$, that is, an element $m = (m_1,\dots,m_k)$ of $\caM_n$ is a $k$-tuple (with varying $k$) of integers $m_i \in \{1,\dots,n\}$ such that $\sum_{i=1}^k m_i = n$. The discrete analogue of the stick-breaking construction is that the probability of $m = (m_1,\dots,m_k)$ is
\be
\bbP_{\caM_n}(m) = \frac1n \, \frac1{n-m_1} \dots \frac1{n-m_1-\dots-m_{k-1}} = \frac1{n M_1 \dots M_{k-1}},
\ee
where we introduced $M_j = n - \sum_{i=1}^j m_i = \sum_{i>j} m_i$.

The split-merge process for GEM consists in choosing two distinct numbers in $\{1,\dots,n\}$ at random. If they fall in different partition elements, these elements are merged and the combined element takes the place of the leftmost one. If the numbers fall in the same partition element $m_j$, it is split uniformly as $m_j = s+t$ ($s$ can be 0, in which case the partition does not change). The $j$th position is assumed by $s$ with probability $\frac s{m_j}$ and by $t$ with probability $\frac t{m_j}$. The other one (call it $u$) takes the $(j+1)$th position with probability $\frac{u}{M_j+u}$ and moves to the right otherwise, where it takes the $(j+2)$th position with probability $\frac{u}{M_{j+1}+u}$, and moves further to the right otherwise.

\begin{centering}
\bfig
\begin{picture}(0,0)%
\includegraphics{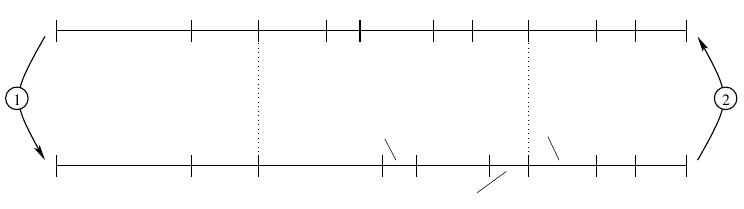}
\end{picture}%
\setlength{\unitlength}{2368sp}%
\begingroup\makeatletter\ifx\SetFigFont\undefined%
\gdef\SetFigFont#1#2#3#4#5{%
  \reset@font\fontsize{#1}{#2pt}%
  \fontfamily{#3}\fontseries{#4}\fontshape{#5}%
  \selectfont}%
\fi\endgroup%
\begin{picture}(9903,2764)(451,-3929)
\put(7576,-2926){\makebox(0,0)[lb]{\smash{{\SetFigFont{7}{8.4}{\rmdefault}{\mddefault}{\updefault}{\color[rgb]{0,0,0}$m_{j+\ell}'=m_{j+\ell+1}$}%
}}}}
\put(4051,-3646){\makebox(0,0)[lb]{\smash{{\SetFigFont{7}{8.4}{\rmdefault}{\mddefault}{\updefault}{\color[rgb]{0,0,0}$m_j'=m_j+m_{j+\ell}$}%
}}}}
\put(9001,-3631){\makebox(0,0)[lb]{\smash{{\SetFigFont{7}{8.4}{\rmdefault}{\mddefault}{\updefault}{\color[rgb]{0,0,0}$m_{k+1}'=m_k$}%
}}}}
\put(5386,-2941){\makebox(0,0)[lb]{\smash{{\SetFigFont{7}{8.4}{\rmdefault}{\mddefault}{\updefault}{\color[rgb]{0,0,0}$m_{j+1}'=m_{j+1}$}%
}}}}
\put(6586,-3871){\makebox(0,0)[lb]{\smash{{\SetFigFont{7}{8.4}{\rmdefault}{\mddefault}{\updefault}{\color[rgb]{0,0,0}$m_{j+\ell-1}'=m_{j+\ell-1}$}%
}}}}
\put(3226,-1441){\makebox(0,0)[lb]{\smash{{\SetFigFont{7}{8.4}{\rmdefault}{\mddefault}{\updefault}{\color[rgb]{0,0,0}$m_2$}%
}}}}
\put(4141,-1426){\makebox(0,0)[lb]{\smash{{\SetFigFont{7}{8.4}{\rmdefault}{\mddefault}{\updefault}{\color[rgb]{0,0,0}$m_j$}%
}}}}
\put(6856,-1456){\makebox(0,0)[lb]{\smash{{\SetFigFont{7}{8.4}{\rmdefault}{\mddefault}{\updefault}{\color[rgb]{0,0,0}$m_{j+\ell}$}%
}}}}
\put(451,-3436){\makebox(0,0)[lb]{\smash{{\SetFigFont{11}{13.2}{\rmdefault}{\mddefault}{\updefault}{\color[rgb]{0,0,0}$m'$}%
}}}}
\put(451,-1636){\makebox(0,0)[lb]{\smash{{\SetFigFont{11}{13.2}{\rmdefault}{\mddefault}{\updefault}{\color[rgb]{0,0,0}$m$}%
}}}}
\put(1606,-1456){\makebox(0,0)[lb]{\smash{{\SetFigFont{7}{8.4}{\rmdefault}{\mddefault}{\updefault}{\color[rgb]{0,0,0}$m_1$}%
}}}}
\put(4786,-1321){\makebox(0,0)[lb]{\smash{{\SetFigFont{7}{8.4}{\rmdefault}{\mddefault}{\updefault}{\color[rgb]{0,0,0}$m_{j+1}$}%
}}}}
\put(7606,-1471){\makebox(0,0)[lb]{\smash{{\SetFigFont{7}{8.4}{\rmdefault}{\mddefault}{\updefault}{\color[rgb]{0,0,0}$m_{j+\ell+1}$}%
}}}}
\put(9031,-1471){\makebox(0,0)[lb]{\smash{{\SetFigFont{7}{8.4}{\rmdefault}{\mddefault}{\updefault}{\color[rgb]{0,0,0}$m_k$}%
}}}}
\put(1471,-3631){\makebox(0,0)[lb]{\smash{{\SetFigFont{7}{8.4}{\rmdefault}{\mddefault}{\updefault}{\color[rgb]{0,0,0}$m_1'=m_1$}%
}}}}
\put(2911,-3646){\makebox(0,0)[lb]{\smash{{\SetFigFont{7}{8.4}{\rmdefault}{\mddefault}{\updefault}{\color[rgb]{0,0,0}$m_2'=m_2$}%
}}}}
\end{picture}%
\caption{The stochastic process on unordered partitions.}
\label{fig partitions}
\efig
\end{centering}

Let $m$ and $m'$ be partitions as in Fig.\ \ref{fig partitions}. $m'$ is obtained from $m$ by merging the elements $m_j$ and $m_{j+\ell}$, which gives $m_j'$; $m$ is obtained from $m'$ by splitting $m_j'$ into $m_j$ and $m_{j+\ell}$ and by placing them in the $j$th and $(j+\ell)$th positions, respectively. The probability of the move $m \mapsto m'$ is
\be
\label{m to m'}
\bbP_{\caM_n}(m) \frac{2 m_j m_{j+\ell}}{n^2}.
\ee
The probability of the move $m' \mapsto m$ is
\be
\label{m' to m}
\begin{split}
&\bbP_{\caM_n}(m') \frac{(m_j')^2}{n^2} \underbrace{\frac{2 m_j}{(m_j')^2}}_{\text{splits $m_j'$}} 
\underbrace{\frac{M_j'}{M_j' + m_{j+\ell}} \frac{M_{j+1}'}{M_{j+1}' + m_{j+\ell}} \dots \frac{M_{j+\ell-2}'}{M_{j+\ell-2}' + m_{j+\ell}}}_{\text{moves $m_{j+\ell}$ by $\ell-1$ steps to the right}} \underbrace{\frac{m_{j+\ell}}{M_{j+\ell-1}' + m_{j+\ell}}}_{\text{stays at position $j+\ell$}}\\
&= \bbP_{\caM_n}(m') \frac{2m_j}{n^2} \frac{M_j'}{M_j} \frac{M_{j+1}'}{M_{j+1}} \dots \frac{M_{j+\ell-2}'}{M_{j+\ell-2}} \frac{m_{j+\ell}}{M_{j+\ell-1}}.
\end{split}
\ee
The expressions \eqref{m to m'} and \eqref{m' to m} are equal, so the probability distribution $\bbP_{\caM_n}$ satisfies the detailed balanced condition and is then invariant.

\section{Relevance of the split-merge process for loop soups}
\label{sec heuristics}

We consider now the model of random loops of Subsection \ref{sec Heisenberg}, but the present heuristics applies to all models that involve macroscopic loops. Let us discretise the ``time" interval $[0,\beta]$ with mesh $1/n$. Given a realisation $\omega$ of crosses and double bars, let $C(\omega)$ and $B(\omega)$ denote the number of crosses and double bars, respectively. The measure on realisations is
\be
\mu(\omega) = \frac1Z \theta^{|\caL(\omega)|} \bigl( \tfrac un \bigr)^{C(\omega)} \bigl( \frac{1-u}n \bigr)^{B(\omega)} \bigl( 1 - \tfrac1n \bigr)^{d |\Lambda| \beta n - c(\omega) - B(\omega)}. 
\ee
Here, $\theta$ is an arbitrary parameter. It needs to be half-integer in order to represent a quantum spin system, but the loop model makes sense more generally.

We now introduce a Markov process such that the measure above is invariant. With $R(\omega,\omega')$ the transition matrix $\omega \mapsto \omega'$, the detailed balance equation is
\be
\theta^{|\caL(\omega)|} \bigl( \tfrac un \bigr)^{C(\omega)} \bigl( \frac{1-u}n \bigr)^{B(\omega)} R(\omega,\omega') = \theta^{|\caL(\omega')|} \bigl( \tfrac un \bigr)^{C(\omega')} \bigl( \frac{1-u}n \bigr)^{B(\omega')} R(\omega',\omega).
\ee
Here is a natural process that satisfies the equation above:
\begin{itemize}
\item A new cross appears in $\{x,y\} \times [t,t+\frac1n]$ at rate $\sqrt\theta \frac un$ if it causes a loop to split; at rate $\frac1{\sqrt\theta} \frac un$ if it causes two loops to merge; at rate $\frac un$ if the number of loops does not change.
\item Same with double bars, but with $1-u$ instead of $u$.
\item An existing cross or double bar is removed at rate $\sqrt\theta$ if its removal causes a loop to split; at rate $\frac1{\sqrt\theta}$ if its removal causes two loops to merge; at rate 1 if the number of loop remains contant.
\end{itemize}
Notice that any new cross or double bar between two loops causes them to merge. When $u=1$, any new cross within a loop causes it to split. When $u=0$, any new double bar within a loop causes it to split, provided the graph $\Lambda$ is bipartite. (We discuss below the case $u \in (0,1)$, where this is not true.)

Let $\gamma, \gamma'$ be two macroscopic loops of lengths $\ell(\gamma), \ell(\gamma')$. They are spread all over $\Lambda$ and they interact between one another, and among themselves, in an essentially mean-field fashion. There exists a constant $c_{1}$ such that a new cross or double bar that causes $\gamma$ to split, appears at rate $\tfrac12 \sqrt\theta \, c_{1} \frac{\ell(\gamma)^{2}}{\beta |\Lambda|}$; a new cross or double bar that causes $\gamma$ and $\gamma'$ to merge appears at rate $(c_{1} / \sqrt\theta) \frac{\ell(\gamma) \ell(\gamma')}{\beta |\Lambda|}$. There exists another constant $c_{2}$ such that the rate for an existing cross or double bar to disappear is $\tfrac12 \sqrt\theta \, c_{2}  \frac{\ell(\gamma)^{2}}{\beta |\Lambda|}$ if $\gamma$ is split, and $(c_{2} / \sqrt\theta) \frac{\ell(\gamma) \ell(\gamma')}{\beta |\Lambda|}$ if $\gamma$ and $\gamma'$ are merged. Consequently, $\gamma$ splits at rate
\be
\tfrac12 \sqrt\theta (c_{1}+c_{2}) \frac{\ell(\gamma)^{2}}{\beta |\Lambda|} \equiv \tfrac12 r_{\rm s} \ell(\gamma)^{2}
\ee
and $\gamma, \gamma'$ merge at rate
\be
\frac1{\sqrt\theta} (c_{1}+c_{2}) \frac{\ell(\gamma) \ell(\gamma')}{\beta |\Lambda|} \equiv r_{\rm m} \ell(\gamma) \ell(\gamma').
\ee
Because of effective averaging over the whole domain, the constants $c_{1}$ and $c_{2}$ are the same for all loops and for both the split and merge events. This key property is certainly not obvious and the interested reader is referred to a detailed discussion for lattice permutations with numerical checks \cite{GLU}. It follows that the lengths of macroscopic loops satisfy an effective split-merge process, and the invariant distribution is Poisson-Dirichlet with parameter $\vartheta = r_{\rm s} / r_{\rm m} = \theta$ \cite{Tsi,Ber,GUW}.

The case $u \in (0,1)$ is different because loops split with only half of the above rate. Indeed, the appearance of a new transition within the loop may just rearrange it: topologically, this is like $0 \leftrightarrow 8$, see Fig.\ \ref{fig 0-8} for illustration. We get Poisson-Dirichlet with parameter $\vartheta = \frac\theta2$.
\bfig
\centerline{\includegraphics[width=70mm]{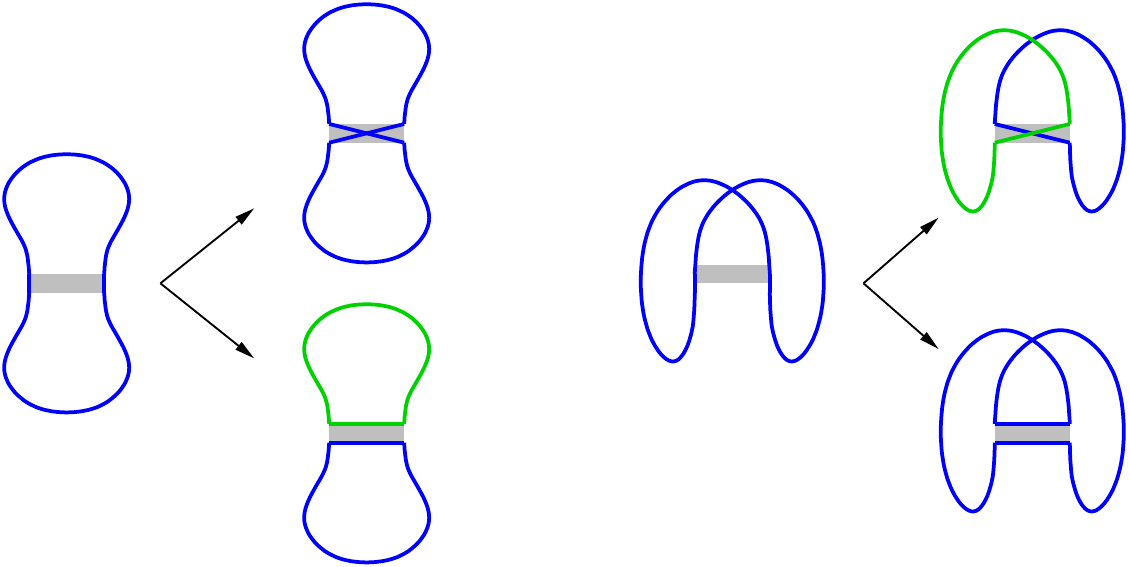}}
\caption{When $u \in (0,1)$, a local change involving two legs of the same loop may rearrange it rather than split it. This figure shows all cases corresponding to the addition of a transition. The loop necessarily splits when $u=0$ or $u=1$.}
\label{fig 0-8}
\efig

\section{Consequences of the Poisson-Dirichlet conjecture}
\label{sec conseq}

Now that we know the structure of the macroscopic loops, we should gain useful information about the original systems. But not too many useful consequences have so far been unearthed. We discuss here quantum spin systems and the symmetry of the low-temperature phases, following \cite{Uel2}.

In this section, we denote $\langle \cdot \rangle_{H_\Lambda}$ the Gibbs state in domain $\Lambda \subset \bbZ^d$ and hamiltonian $H_\Lambda$, that is,
\be
\langle \cdot \rangle_{H_\Lambda} = \frac1{\Tr \e{-\beta H_\Lambda}} \Tr \cdot \e{-\beta H_\Lambda}.
\ee

\subsection{Spin $\frac12$ systems}

We consider the hamiltonian of Eq.\ \eqref{ham spin 12} with nearest-neighbour interactions $-S_x^1 S_y^1 - (2u-1) S_x^2 S_y^2 - S_x^3 S_y^3$. In dimensions 3 and larger (or in the ground state in dimension 2), one expects ferromagnetism and long-range order. The case $u=1$ corresponds to the ordinary Heisenberg ferromagnet and the extremal states should be given by
\be
\langle \cdot \rangle_{\vec a} = \lim_{h \to 0+} \lim_{\Lambda \nearrow \bbZ^d} \langle \cdot \rangle_{H_\Lambda^{(u)} - h \sum_x \vec a \cdot \vec S_x},
\ee
where $\vec a$ is any vector in $\bbS^2$. In the case where $u \in (0,1)$, the model has U(1) symmetry only, and the extremal states are $\langle \cdot \rangle_{\vec a}$ with $\vec a \in \bbS^1$, of the form $(a_1,0,a_3)$. Another way to write the symmetry breakings is
\be
\label{Spencer}
\lim_{\Lambda \nearrow \bbZ^d} \langle \e{\frac h{|\Lambda|} \sum_{x\in \Lambda} \vec a \cdot \vec S_x} \rangle_{H_\Lambda^{(u)}} = \begin{cases} \int_{\bbS^2} \e{h \frm \vec a \cdot \vec b} \dd\vec b & \text{if $u=0$ or $1$}, \\ \int_{\bbS^1} \e{h \frm \vec a \cdot \vec b} \dd\vec b & \text{if } u \in (0,1). \end{cases}
\ee
Here, $\frm$ is the magnetisation of the system\footnote{Tom Spencer suggested these equations (private communication).}. In the case $u \in (0,1)$, both $\vec a$ and $\vec b$ are of the form $(a_1,0,a_3)$.
The meaning of these identities is that the infinite volume limit of $\langle \cdot \rangle_{H_\Lambda}$ is a convex combination of the states $\langle \cdot \rangle_{\vec a}$ above.
By rotation invariance, this does not depend on $\vec a$ and we have for $u=0$ or 1,
\be
\label{Spencer 1}
\int_{\bbS^2} \e{h \frm \vec a \cdot \vec b} \dd\vec b = \int_{\bbS^2} \e{h \frm b_3} \dd\vec b = \frac{\sinh(h \frm)}{h \frm}.
\ee
In the case $u \in (0,1)$, we get a Bessel function, namely,
\be
\label{Spencer 2}
\int_{\bbS^1} \e{h \frm \vec a \cdot \vec b} \dd\vec b = \frac1{2\pi} \int_0^{2\pi} \e{h \frm \cos\phi} \dd\phi = \sum_{n\geq0} \frac1{(n!)^2} (\tfrac12 h\frm)^{2n}.
\ee

The advantage of the identities \eqref{Spencer} is that the expectation of $\e{\frac h{|\Lambda|} \sum_x S_x^3}$ has a nice expression in terms of the loops of Section \ref{sec Heisenberg}. Indeed, by a similar expansion that uses Trotter product formula, we get
\be
\langle \e{\frac h{|\Lambda|} \sum_x S_x^3} \rangle_{H_\Lambda^{(u)}} = \bbE_\Lambda^{(u)} \Bigl( \prod_{\gamma \in \caL(\omega)} \cosh \Bigl( \frac h{2|\Lambda|} \ell(\gamma) \Bigr) \Bigr).
\ee
Here, $\bbE_\Lambda^{(u)}$ denotes expectation with respect to the model of random loops with weights $2^{|\caL(\omega)|}$, and $\ell(\gamma)$ is the total length of all vertical legs of the loop $\gamma$.

At low temperatures and for large domains, we expect that macroscopic loops are present and that they occupy a fixed fraction $m$ of the available space. Further, by the discussion of Section \ref{sec heuristics}, the joint distribution of their lengths is Poisson-Dirichlet with parameter $\vartheta=2$ when $u=0$ or 1, and $\vartheta=1$ when $u \in (0,1)$. By Conjecture 2 (3), which applies to the hyperbolic cosine, we get
\be
\lim_{|\Lambda|\to\infty} \langle \e{\frac h{|\Lambda|} \sum_x S_x^3} \rangle_{H_\Lambda^{(u)}} = \bbE_{{\rm PD}(\vartheta)} \Bigl( \prod_{j\geq1} \cosh \bigl(\tfrac12 hm X_j \bigr) \Bigr).
\ee
The right side was calculated in Eq.\ \eqref{calc cosh}; we obtained
\be
\label{Spencer PD}
\bbE_{{\rm PD}(\vartheta)} \Bigl( \prod_{j\geq1} \cosh \bigl( \tfrac12 hm X_j \bigr) \Bigr) = \begin{cases} \sum_{r\geq0} \frac1{(2r+1)!} (\frac{hm}2)^{2r} = \frac{\sinh(\frac12 hm)}{\frac12 hm} & \text{if } \vartheta=2, \\ \sum_{r\geq0} \frac1{(r!)^2} (\frac14 hm)^{2r} & \text{if } \vartheta=1. \end{cases}
\ee
(The last expression is perhaps not immediately apparent from \eqref{calc cosh}; it uses the identity $2^{2n} n! \Gamma(n+\frac12) = \Gamma(2n+1) \Gamma(\frac12)$.) Then Eqs \eqref{Spencer PD} are precisely the expressions \eqref{Spencer 1} and \eqref{Spencer 2}, with the magnetisation being half the mass of macroscopic loops, $\frm = \frac12 m$. This shows that the Poisson-Dirichlet conjectures are compatible with our expectations of symmetry breaking.

\subsection{Spin 1 systems}

We now turn to the spin 1 model of Eq. \eqref{ham spin 1}; it is worth to consider the general model with SU(2) invariant, nearest-neighbour interactions, namely
\be
\label{gen ham spin 1}
H_\Lambda = -\sumtwo{\{x,y\} \subset \Lambda}{\|x-y\|=1} \bigl( J_1 \vec S_x \cdot \vec S_y + J_2 (\vec S_x \cdot \vec S_y)^2 \bigr).
\ee
Here, $J_1,J_2$ are two real parameters. The phase diagram of this model was studied in \cite{FKK}. For $d\geq3$ and low temperatures (or $d=2$ in the ground state), it decomposes into four regions with ferromagnetic, spin nematic, antiferromagnetic, and staggered nematic phases. The phase diagram is displayed in Fig.\ \ref{fig phd}.

\bfig
\includegraphics[width=8cm]{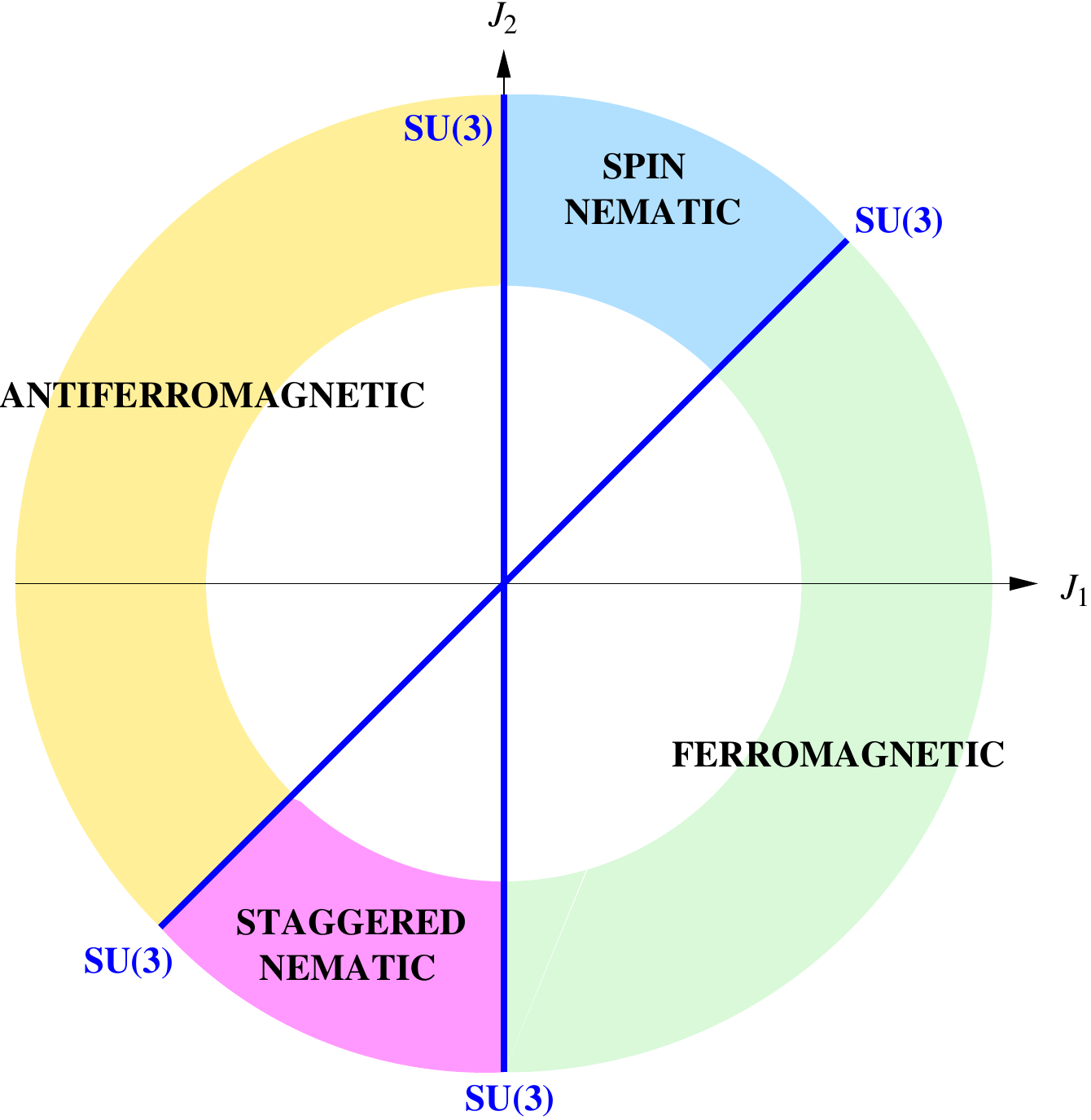}
\caption{Phase diagram of the general spin 1 model with hamiltonian \eqref{gen ham spin 1} in dimension $d\geq3$. On the two lines $J_{1}=0$ and $J_{2}=J_{1}$ the model has SU(3) invariance, not only SU(2). The phase diagram is expected to show four phases (ferromagnetic, nematic, antiferromagnetic, staggered nematic) that are separated by those lines.}
\label{fig phd}
\efig

The loop representation of Section \ref{sec Heisenberg} applies to the model with the hamiltonian $\tilde H_\Lambda^{(u)}$ in \eqref{ham spin 1}, which corresponds to the spin nematic region, and also to its boundaries where the model has SU(3) invariance. We only discuss the case $u \in (0,1)$.

We now seek to confront symmetry breaking with the Poisson-Dirichlet conjectures in a similar fashion as in the spin $\frac12$ case. This is actually more interesting here because the nature of symmetry breaking is no longer obvious. The operators that are associated with the spin nematic phase are
\be
A^{\vec a}_x = (\vec a \cdot \vec S_x)^2 - \tfrac23,
\ee
with $\vec a \in \bbS^2$ (notice that $\vec a$ is equivalent to $-\vec a$). The constant $-\frac23$ ensures that $\langle A_x^{\vec a} \rangle = 0$ when the Gibbs state is invariant under spin rotations. We write $A_x^i$ for $A_x^{\vec e_i}$.

We first look for an analogue to the identities \eqref{Spencer}. Assuming that a spin nematic transition takes place, there exist extremal Gibbs states $\langle \cdot \rangle_{\vec a}$ where $\vec a \in \bbS^2$, and with $\langle \cdot \rangle_{-\vec a} = \langle \cdot \rangle_{\vec a}$. (It might be more elegant to label extremal states with the projective space P$\bbS^2$, where $\pm \vec a$ are identified.) We introduce what should be the nematic counterpart to the magnetisation density, namely
\be
\frn = \lim_{\Lambda \nearrow \bbZ^d} \Bigl\langle \frac1{|\Lambda|} \sum_{x \in \Lambda} A_x^3 \Bigr\rangle_{\vec e_3}.
\ee
We expect that $\frn \neq 0$ if the temperature is low and $d\geq3$, or in the ground state and $d\geq2$. The expectation of $\frac1{|\Lambda|} \sum_x A_x^{\vec a}$ for general $\vec a \in \bbS^2$ can be expressed in terms of $\frn$. Indeed,
\be
\Bigl\langle \frac1{|\Lambda|} \sum_{x \in \Lambda} A_x^{\vec a} \Bigr\rangle_{\vec e_3} = \frac1{|\Lambda|} \Bigl[ \sum_{i=1}^3 a_i^2 \langle (S_x^i)^2 - \tfrac23 \rangle_{\vec e_3} + \sum_{i\neq j} a_i a_j \langle S_x^i S_x^j \rangle_{\vec e_3} \Bigr].
\ee
We can assume that $\langle \cdot \rangle_{\vec e_3}$ is invariant under spin rotations around $\vec e_3$, and also that $\langle S_x^3 \rangle_{\vec e_3} = 0$, so that $\langle S_x^i S_x^j \rangle_{\vec e_3} = 0$ for all $i \neq j$. Further, since $(S_x^1)^2 + (S_x^2)^2 + (S_x^3)^2 = 2$, we have
\be
\langle (S_x^1)^2 - \tfrac23 \rangle_{\vec e_3} =  \langle (S_x^2)^2 - \tfrac23 \rangle_{\vec e_3} = -\tfrac12  \langle (S_x^3)^2 - \tfrac23 \rangle_{\vec e_3}.
 \ee
This gives
\be
\lim_{\Lambda \nearrow \bbZ^d} \Bigl\langle \frac1{|\Lambda|} \sum_{x\in\Lambda} A_x^{\vec a} \Bigr\rangle_{\vec e_3} = \frn (a_3^2 - \tfrac12 a_1^2 - \tfrac12 a_2^2).
\ee
This allows to calculate
\be
\label{premiere expression}
\begin{split}
\lim_{\Lambda \nearrow \bbZ^d} \Bigl\langle \e{\frac h{|\Lambda|} \sum_{x\in\Lambda} A_x^3} \Bigr\rangle_{\tilde H_\Lambda^{(u)}} &= \lim_{\Lambda \nearrow \bbZ^d} \int_{\bbS^2} \Bigl\langle \e{\frac h{|\Lambda|} \sum_{x\in\Lambda} A_x^3} \Bigr\rangle_{\vec a} \dd\vec a \\
&=  \lim_{\Lambda \nearrow \bbZ^d} \int_{\bbS^2} \Bigl\langle \e{\frac h{|\Lambda|} \sum_{x\in\Lambda} A_x^{\vec a}} \Bigr\rangle_{\vec e_3} \dd\vec a \\
&=  \int_{\bbS^2} \e{h \frn (a_3^2 - \frac12 a_1^2 - \frac12 a_2^2)} \dd\vec a \\
&= \e{-\frac12 h\frn} \sum_{k\geq0} \frac{(\frac32 h\frn)^k}{k! (2k+1)}.
\end{split}
\ee

Next, we compute the same quantity using the loop representation and the conjectures. By a Trotter product expansion, we obtain
\be
\Bigl\langle \e{\frac h{|\Lambda|} \sum_{x\in\Lambda} A_x^3} \Bigr\rangle_{\tilde H_\Lambda^{(u)}} = \bbE_\Lambda^{(u)} \Bigl( \prod_{\gamma \in \caL(\omega)} \bigl( \tfrac13 \e{-\frac23 \frac h{|\Lambda|} \ell(\gamma)} + \tfrac23 \e{\frac13 \frac h{|\Lambda|} \ell(\gamma)} \bigr) \Bigr),
\ee
where the expectation is taken over the random loop model of Section \ref{sec Heisenberg} with weights $3^{|\caL(\omega)|}$. Conjectures 1 and 2, together with the argument of Section \ref{sec heuristics}, state that macroscopic loops have fixed total mass $m$, and that the joint distribution of their lengths is Poisson-Dirichlet with parameter $\vartheta=\frac32$. By Conjecture 2 (3), we have
\be
\begin{split}
\lim_{\Lambda \nearrow \bbZ^d} \Bigl\langle \e{\frac h{|\Lambda|} \sum_{x\in\Lambda} A_x^3} \Bigr\rangle_{\tilde H_\Lambda^{(u)}} &= \bbE_{{\rm PD}(\frac32)} \Bigl( \prod_{i\geq1} \bigl( \tfrac13 \e{-\frac23 hm Y_i} + \tfrac23 \e{\frac13 hm Y_i} \bigr) \Bigr) \\
&=  \e{-\frac23 hm} \bbE_{{\rm PD}(\frac32)} \Bigl( \prod_{i\geq1} \bigl( \tfrac13 + \tfrac23 \e{hm Y_i} \bigr) \Bigr).
\end{split}
\ee
We can use Eq. \eqref{ca aussi !} for the function $f(s) = \frac13 + \frac23 \e{s}$ whose Taylor coefficients are $a_0=1$, $a_k = \frac23 \frac1{k!}$ for $k\geq1$. We obtain
\be
\begin{split}
\lim_{\Lambda \nearrow \bbZ^d} \Bigl\langle \e{\frac h{|\Lambda|} \sum_{x\in\Lambda} A_x^3} \Bigr\rangle_{\tilde H_\Lambda^{(u)}} &= \e{-\frac23 hm} \sum_{n\geq0} \frac1{n!} \sum_{k_1,\dots,k_n\geq1} a_{k_1} \dots a_{k_n} \frac{(\frac32)^n \Gamma(\frac32) \Gamma(k_1) \dots \Gamma(k_n)}{\Gamma(\frac32 + k_1 + \dots + k_n)} (hm)^{\sum k_i} \\
&= \e{-\frac23 hm} \sum_{r\geq0} \frac{(hm)^r \Gamma(\frac32)}{\Gamma(\frac32+r)} \sum_{n\geq0} \frac1{n!} \sumtwo{k_1,\dots,k_n\geq1}{k_1+\dots+k_n=r} \frac1{k_1 \dots k_n} \\
&= \e{-\frac23 hm} \sum_{r\geq0} \frac{\Gamma(\frac32)}{\Gamma(\frac32+r)} (hm)^r.
\end{split}
\ee
We used Eq.\ \eqref{cool identity} in the last equality. Although it is not immediately apparent, this is the same function of $h$ as \eqref{premiere expression}, provided that
\be
\label{et voila}
\frn = -\tfrac23 m.
\ee

Recall that $m$ represents the fraction of available volume that is occupied by macroscopic loops and it is therefore nonnegative. It may come as a surprise that $\frn$ is negative. This provides information on the nature of the nematic states. Indeed, it is natural to conjecture that extremal nematic states are defined in a similar manner as in the classical case, namely,
\be
\label{axial nematic}
\langle \cdot \rangle_{\vec a} = \lim_{h\to0+} \lim_{\Lambda\nearrow\bbZ^d} \langle \cdot \rangle_{\tilde H_\Lambda^{(u)} - h \sum_{x\in\Lambda} A_x^{\vec a}}.
\ee
These are ``axial nematic" states \cite{FKK}. The state $\langle \cdot \rangle_{\vec e_3}$ has an illuminating expression in terms of random loops. With hamiltonian $\tilde H_\Lambda^{(u)} - h \sum_{x\in\Lambda} A_x^3$, the partition function becomes
\be
Z_{\vec e_3}^{(u)}(\Lambda,h) = \e{\frac13 \beta h |\Lambda|} \int \rho(\dd\omega) \prod_{\gamma \in \caL(\omega)} \sum_{\sigma_\gamma \in \{-1,0,1\}} \e{h \ell(\gamma) (\sigma_\gamma^2-1)}.
\ee
We should keep in mind that the domain $\Lambda$ is huge and the parameter $h$ is small and positive, with $|\Lambda|^{-1} \ll h \ll 1$. It follows that short loops carry labels $\{-1,0,1\}$ indifferently, while macroscopic loops carry labels $\{-1,1\}$. (These labels are not exactly constant along each loop, but they change signs when the vertical direction changes.) The weight is therefore $3^{|\caL_{\rm short}(\omega)|} 2^{|\caL_{\rm long}(\omega)|}$; let $\bbP_{\vec e_3}^{(u)}$ denote the corresponding loop measure. This allows to relate $\frn$ and $m$:
\be
\begin{split}
\langle A_x^3 \rangle_{\vec e_3} &= \frac{\e{\frac13 \beta h |\Lambda|}}{Z_{\vec e_3}(\Lambda,h)} \biggl[ \int_{\gamma_{(x,0) \text{ is short}}} \rho(\dd\omega) \; 3^{|\caL_{\rm short}(\omega)|} 2^{|\caL_{\rm long}(\omega)|} \; \tfrac13 \sum_{\sigma \in \{-1,0,1\}} (\sigma^2 - \tfrac23) \\
&\qquad\qquad + \int_{\gamma_{(x,0) \text{ is long}}} \rho(\dd\omega) \; 3^{|\caL_{\rm short}(\omega)|} 2^{|\caL_{\rm long}(\omega)|} \; \tfrac12 \sum_{\sigma \in \{-1,1\}} (\sigma^2 - \tfrac23) \biggr] \\
&= \tfrac13 \bbP_{\vec e_3}^{(u)}( \gamma_{(x,0) \text{ is long}} ).
\end{split}
\ee
We split the integral over all loop configurations according to whether $(x,0) \in \Lambda \times [0,\beta]$ belongs to a short or long loop. The sums $\sum_\sigma$ are over the spin values of the loop $\gamma_{(x,0)}$.
The latter probability is equal to $m$, which gives $\frn = \frac13 m$. This contradicts \eqref{et voila}, however. Where is the error?

It turns out that the extremal nematic states are not axial nematic, but ``planar nematic" \cite{FKK}. That is, let
\be
\langle \cdot \rangle_{\vec a}' = \lim_{h\to0+} \lim_{\Lambda\nearrow\bbZ^d} \langle \cdot \rangle_{\tilde H_\Lambda^{(u)} + h \sum_{x\in\Lambda} A_x^{\vec a}}.
\ee
Notice the ``$+$" sign in front of $h$, which should be contrasted with Eq.\ \eqref{axial nematic}. This state favours the eigenvalue 0 rather than $\pm 1$. The corresponding partition function is
\be
Z_{\vec e_3}'(\Lambda,h) = Z_{\vec e_3}^{(u)}(\Lambda,-h) = \e{\frac23 \beta h |\Lambda|} \int \rho(\dd\omega) \prod_{\gamma \in \caL(\omega)} \sum_{\sigma_\gamma \in \{-1,0,1\}} \e{-h \ell(\gamma) \sigma_\gamma^2}.
\ee
When $|\Lambda|^{-1} \ll h \ll 1$, the short loops carry labels $\{-1,0,1\}$ as before, but long loops are stuck with label 0. Then, with $\bbP_{\vec e_3}'$ denoting the corresponding loop measure,
\be
\begin{split}
\langle A_x^3 \rangle_{\vec e_3}' &= \frac{\e{\frac23 \beta h |\Lambda|}}{Z_{\vec e_3}'(\Lambda,h)} \biggl[ \int_{\gamma_{(x,0) \text{ is short}}} \rho(\dd\omega) \; 3^{|\caL_{\rm short}(\omega)|} \; \tfrac13 \sum_{\sigma \in \{-1,0,1\}} (\sigma^2 - \tfrac23) \\
&\qquad\qquad + \int_{\gamma_{(x,0) \text{ is long}}} \rho(\dd\omega) \; 3^{|\caL_{\rm short}(\omega)|} (-\tfrac23) \biggr] \\
&= -\tfrac23 \bbP_{\vec e_3}'( \gamma_{(x,0) \text{ is long}} ).
\end{split}
\ee
This gives $\frn = -\frac23 m$, in conformity with \eqref{et voila}. These calculations use the conjectures about the joint distribution of lengths of long loops, and they give strong evidence that nematic states are planar nematic. This result was far from immediate.

Similar considerations are possible in the cases $u=0$ and $u=1$, which correspond to SU(3)-invariant interactions. We refer to \cite{Uel2} for details.

\bigskip
\noindent
{\bf Acknowledgments:} I am grateful to Bogdan Cichocki, Filip Dutka, Pawe\l{} Jakubczyk, Maciej Lisicki, Andrzej Majhofer, Marek Napi\'orkowski, Jaros\l aw Piasecki, and Piotr Szymczak, who organised the 6th Warsaw School of Statistical Physics, and who gave me the opportunity to give a series of lectures on one of my favourite topics. Pawe\l{} Jakubczyk and Marcin Napi\'orkowski made useful suggestions on these notes. I would also like to thank the many colleagues and collaborators who helped me to understand this topic better, including J\"urg Fr\"ohlich, Gian Michele Graf, Alan Hammond, and James Martin. I also acknowledge support from The Leverhulme Trust through the International Network `Laplacians, Random Walks, Quantum Spin Systems'.

\renewcommand{\refname}{\small References}
\bibliographystyle{symposium}

\end{document}